\documentclass[sigconf,10pt]{acmart}
\usepackage{algorithm}
\usepackage{algpseudocode}
\usepackage{amsmath}
\usepackage{stfloats}
\usepackage[export]{adjustbox}
\usepackage{booktabs}
\usepackage{tabularx}
\usepackage{graphicx}
\usepackage[justification=centering]{caption}
\usepackage{subcaption}
\usepackage{listings}
\usepackage{url}
\usepackage{hyperref}
\usepackage{xcolor}
\usepackage{circledsteps}
\usepackage[normalem]{ulem}

\usepackage{booktabs, multirow}
\usepackage{soul}
\usepackage{changepage,threeparttable}

\copyrightyear{2026}
\acmYear{2026}
\setcopyright{cc}
\setcctype{by}
\acmConference[SYSTOR '26]{The 19th ACM International Systems and Storage Conference}{September 08--09, 2026}{Virtual Event, Israel}
\acmBooktitle{The 19th ACM International Systems and Storage Conference (SYSTOR '26), September 08--09, 2026, Virtual Event, Israel}
\acmDOI{10.1145/3793230.3837768}
\acmISBN{979-8-4007-2473-2/2026/09}

\newcommand{\EK}[1]{}
\newcommand{\capPoTwoEst}{31.6}
\newcommand{\capPoTwoOracle}{32.4}
\newcommand{\capFanoutEst}{31.7}
\newcommand{\capFanoutOracle}{32.6}
\newcommand{\capLlumnix}{31.5}
\newcommand{\capPoFourEst}{31.9}
\newcommand{\capPoEightEst}{31.7}
\newcommand{\capQwenPoTwoEst}{73.9}
\newcommand{\capQwenPoTwoOracle}{78.5}
\newcommand{\capQwenLlumnix}{69.7}
\newcommand{\capBurstGPTPoTwoOracle}{64.5}
\newcommand{\sloSec}{10}
\newcommand{\cpuPoTwoMeanAtTwenty}{20.3}

\newcommand{\cpuFanoutMeanAtTwenty}{56.3}
\newcommand{\cpuPoTwoVsFanoutRatio}{2.8}

\begin{document}
\raggedbottom
\title{Astrolabe: Balancing Load in LLM Serving with Randomized Prediction-Guided Scheduling}

\author{Wei Da}
\authornote{Corresponding author.}
\orcid{0009-0003-7305-6531}
\email{wd312@cam.ac.uk}
\affiliation{%
  \institution{University of Cambridge}
  \city{Cambridge}
  \country{United Kingdom}}

\author{Evangelia Kalyvianaki}
\orcid{0000-0003-0753-1261}
\email{ek264@cam.ac.uk}
\affiliation{%
  \institution{University of Cambridge}
  \city{Cambridge}
  \country{United Kingdom}}

\begin{abstract}
This paper presents Astrolabe, a \emph{randomized} prediction-guided scheduler for one-shot request dispatch in multi-instance large language model (LLM) serving. Astrolabe improves load balancing without relying on migration-based rebalancing, whose KV-cache transfers can introduce substantial overhead and network contention under high load. Astrolabe combines response-length estimation, per-instance simulation-based latency prediction, and a \emph{power-of-two choices} dispatch policy to improve load balancing and avoid request herding. On the default Llama-2-7B/ShareGPT setup, Astrolabe matches the best load-aware baseline's SLO capacity (31.6 vs 31.5 QPS) while improving latency metrics: 8--36\% lower mean time-to-first-token (TTFT), 16--77\% lower P99 TTFT, up to 5.6\% lower mean end-to-end (E2E) latency, and $\sim$6$\times$ fewer preemptions once capacity is reached. Under configuration shifts, Astrolabe also improves SLO capacity by up to +6\% on Qwen2-7B and +7.1\% under tight batching, with 6--9\% lower mean E2E under bursty arrivals, while achieving a $\sim$\cpuPoTwoVsFanoutRatio{}-fold reduction in per-predictor CPU relative to full fanout. With migration enabled on A100, Astrolabe outperforms Llumnix by up to 2.6$\times$ in throughput with orders-of-magnitude lower per-token latency at saturation.
\end{abstract}


\maketitle

\section{Introduction} \label{section:introduction}

The rise of large language models like GPT~\cite{achiam2023gpt}, Llama~\cite{llama}, Gemini~\cite{team2023gemini}, Qwen~\cite{yang2025qwen3}, and DeepSeek~\cite{liu2025deepseek} has revolutionized modern applications, such as chatbots~\cite{achiam2023gpt}, virtual assistants~\cite{guan2023intelligentvirtualassistantsllmbased}, code generation~\cite{jiang2024surveylargelanguagemodels}, and creative writing~\cite{gómezrodríguez2023confederacymodelscomprehensiveevaluation}. This places immense pressure on LLM inference serving systems, which must meet stringent latency requirements for a seamless user experience and maximize throughput to handle growing demand~\cite{shen2025accelgenheterogeneoussloguaranteedhighthroughput, cheng2025scootsloorientedperformancetuning, vllm, orca}. To this end, recent years have seen significant progress in LLM serving systems. Several key techniques, such as continuous batching~\cite{orca}, Paged Attention~\cite{vllm}, chunked prefill~\cite{agrawal2023sarathiefficientllminference}, and FlashAttention~\cite{flashattentionv2}, have been proposed to improve throughput, latency, and utilization.

However, despite the above advances, LLM inference is often characterized as unpredictable~\cite{sun2024llumnix}. For example, LLMs generate tokens autoregressively based on preceding tokens until a stop signal is reached, leading to variable response lengths and numbers of decoding steps~\cite{schuurmans2024autoregressivelargelanguagemodels}.
Paged Attention~\cite{vllm}, which dynamically allocates memory resources and allows request preemption, further contributes to the dynamic nature of runtime memory consumption. Decode-step latency also exhibits high variance due to batch-size variation and interference between requests in the same batch~\cite{sun2024llumnix}.

These uncertainties make load balancing across LLM serving instances challenging: common runtime metrics (latency, throughput, memory) no longer represent each request's E2E load. Moreover, unexpectedly long responses further load already-overloaded hosts and block subsequent requests (head-of-line blocking~\cite{QLMQueue, ELIS}). In practice, production multi-instance frameworks~\cite{nvidiaTritonInference, deepspeed} dispatch with simple heuristics like round-robin, offering no performance guarantee.  Alternative solutions, such as Llumnix~\cite{sun2024llumnix}, combine dispatching with dynamic rebalancing through live migration and have demonstrated performance improvements. However, migration requires transferring request KV caches~\cite{kang2024gearefficientkvcache} across instances, adding network, memory, and orchestration overhead. Under high load, these costs can accumulate and contend with normal serving traffic, potentially degrading tail latency and throughput.

In parallel, a line of work mitigates such uncertainties with purpose-built assistant models.
For example, response lengths can be predicted by a pretrained regression model~\cite{length_estimation} and execution latency by an inference simulator~\cite{vidur}.  This presents an unexplored opportunity: if length and latency can be estimated at dispatch time, the scheduler can proactively route long requests to less-loaded instances before imbalance manifests.

A basic way to exploit such predictions is to query every instance's predictor in parallel and pick the lowest-latency candidate (\emph{full fanout}). This gives the best load-balancing quality but has three practical costs. \emph{First}, each dispatch costs $O(N)$ predictor queries for $N$ instances, which scales poorly in the scheduler control plane. \emph{Second}, simulation runs on CPU to predict the scheduling signal (e.g., latency) for each candidate instance, so aggregate cluster-CPU grows linearly with $N$ per request. \emph{Third}, deterministic fanout creates a \emph{thundering-herd} issue: under short correlated bursts, nearby requests compute similar ``best'' candidates and route to the same instance, overloading it transiently~\cite{mitzenmacher2001,10.1137/S0097539795288490}. These costs matter in any large multi-instance deployment, and become especially acute when scheduler CPU, control-plane bandwidth, or interconnect capacity is limited.

In this paper, we present \emph{Astrolabe}, a randomized prediction-guided scheduler for multi-instance LLM inference clusters. Astrolabe's default dispatch policy is \emph{power-of-two choices} (Po2): for each request, sample two random instances, query their predictors in parallel, and route to the lower-predicted-latency candidate. This yields constant per-request fanout cost, a $\sim$\cpuPoTwoVsFanoutRatio{}-fold reduction in per-predictor CPU relative to full fanout in our measurements, and crucially breaks herding by randomizing the probe set so that simultaneous requests are unlikely to sample the same pair of candidates.

Classical power-of-two choices analysis~\cite{mitzenmacher2001} has been widely applied to load balancing in distributed systems~\cite{sparrow, prequal, hawk, dodoor}, and our measurements confirm the effect in the LLM-serving context. Astrolabe is not tied to a particular cluster topology: it makes high-quality placement decisions before execution, avoiding migration where it is unavailable, expensive, or harmful due to KV-cache-transfer overhead and network contention. Under high load, migration overhead can collapse serving performance while Astrolabe sustains higher throughput and lower latency (\S\ref{section:llumnix_comparison}).

Astrolabe's request path is deliberately simple. It first predicts a request's response length from the prompt using a lightweight regression model. The scheduler then samples two candidate instances and queries their Predictor sidecars, each of which simulates the request against its local runtime state and returns an expected latency. The request is dispatched to the candidate with the lower predicted latency. This per-instance view improves balance with one-shot dispatch, avoids live migration, and stays compatible with existing backends; the Predictor interface is pluggable across simulators, frameworks, and scheduling objectives.

We evaluate Astrolabe against five heuristic/load-aware schedulers and alternative prediction-guided variants on a 12-GPU A30 cluster across ShareGPT, BurstGPT, Llama-2-7B, and Qwen2-7B; we also compare against the official Llumnix release on a 2-node A100 cluster with full migration enabled. On the default A30 Llama-2-7B/ShareGPT setup, Astrolabe preserves Llumnix's SLO capacity (\capPoTwoEst{} vs \capLlumnix{} QPS with no migration), matches full-fanout capacity (\capPoTwoEst{} vs \capFanoutEst{} QPS with no sampling) at 6$\times$ fewer predictor queries, and delivers the lowest time-to-first-token (TTFT) and end-to-end (E2E) latency metrics at 15 of 17 tested QPS points. It also shows up to 97\% lower mean TTFT than round-robin, 8--38\% lower than the best baseline, and $\sim$6$\times$ fewer preemptions at overload ($\sim$300 vs $\sim$1{,}800 at QPS=32). Under configuration shifts, Astrolabe lifts capacity by up to +6\% on Qwen2-7B and +7.1\% on tight batching, tolerates injected prediction noise (at most +6.3\% mean E2E and +3.7\% P99 E2E across the heatmap), and reduces mean E2E by 6--9\% under bursty arrivals. On the two-node A100 cluster, Astrolabe outperforms full Llumnix by up to 2.6$\times$ in throughput and achieves orders-of-magnitude lower per-token latency at saturation. These results show that migration overhead can collapse serving performance under high load, and that prediction-guided one-shot dispatch can provide equal or better load balancing without migration.

This paper makes three main contributions. \emph{First}, we explain why multi-instance LLM serving load is hard to infer from runtime metrics and identify \emph{scheduler herding} issues at \S\ref{section:motivation} and \S\ref{section:scheduling_issues}.
\emph{Second}, we design and implement \emph{Astrolabe}, a one-shot randomized prediction-guided scheduler combining response-length estimation, per-instance simulation-based latency prediction, and a power-of-two choices dispatch policy to achieve high-quality load balancing and request scheduling at \S\ref{section:system_design} and \S\ref{section:implementation}.
\emph{Third}, we evaluate Astrolabe on a 12-GPU A30 cluster and a 2-node A100 cluster across two models, two datasets, and four backend configurations, quantifying prediction accuracy, latency, throughput, CPU overhead, robustness (with bursty arrivals or injected noise), and a migration-based-rebalancing case study (\S\ref{section:evaluation}).

\section{Background}\label{section:motivation}
LLMs build on transformer architectures~\cite{transformer}. A prompt is tokenized and processed by transformer blocks to produce intermediate representations (\emph{prefill}); tokens are then generated autoregressively (\emph{decode}) until an end-of-sequence token or max-length limit. Because tokens are reused across steps, attention key/value tensors are cached (\emph{KV cache}~\cite{kang2024gearefficientkvcache}) to avoid recomputation. Prefill is typically compute-bound; decode is memory-bandwidth-bound~\cite{podAttention} and cannot be parallelized across tokens.

Requests are grouped into batches to leverage GPU matrix units; each decoding step generates one token for every in-batch request, and KV cache grows with tokens. Static batching~\cite{lightseqhighperformanceinference, turbotransformersefficientgpuserving} fixes the batch set and under-utilizes slots when shorter requests finish early. Continuous Batching~\cite{orca} addresses this by letting completed requests exit and admitting new successors (FCFS order) up to a maximum batch size. A local scheduler in the inference framework forms these execution batches.

\begin{figure}
    \centering
    \includegraphics[width=1.0\linewidth]{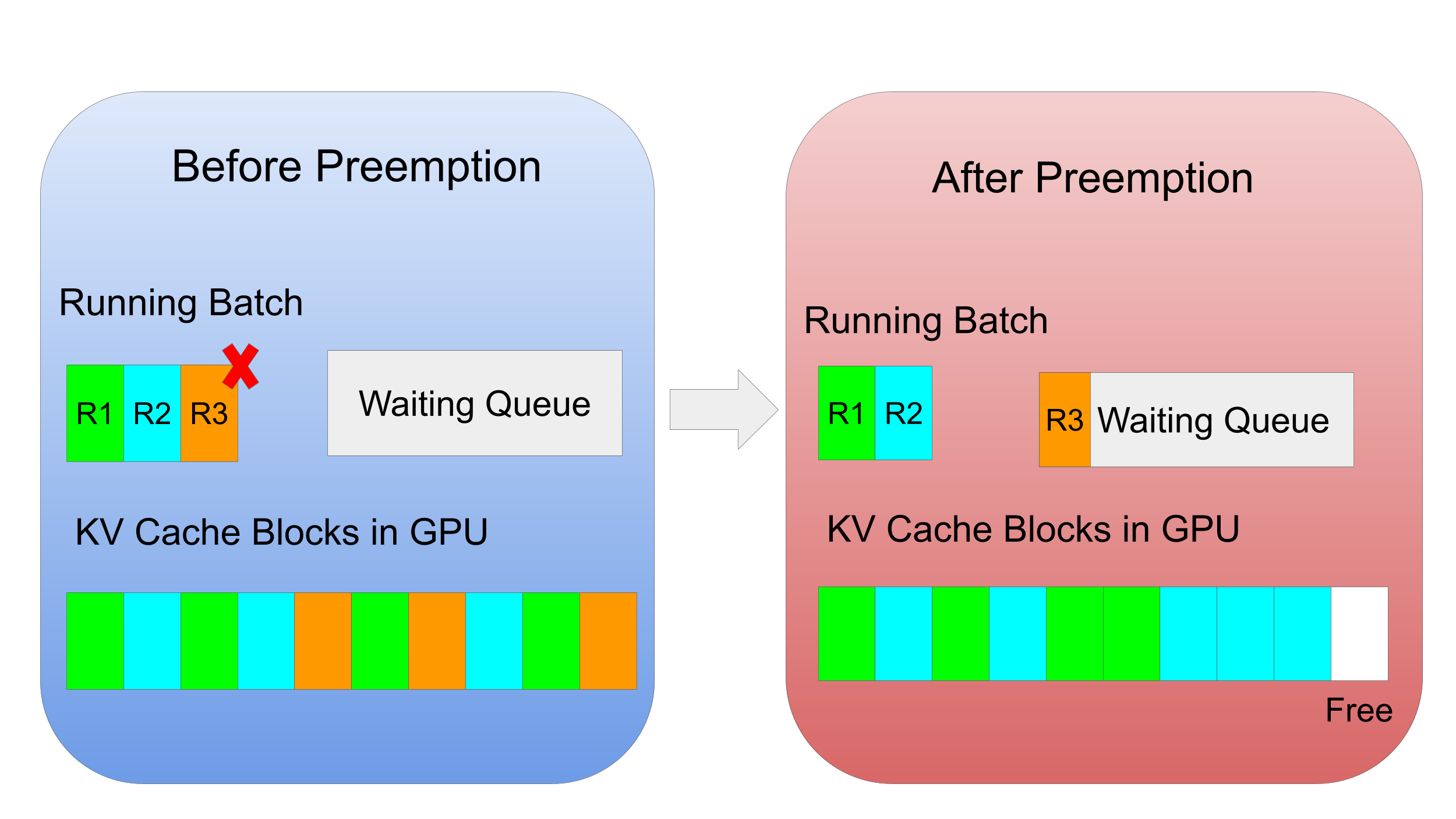}
    \caption{Paged Attention and Preemption}%
    \Description{Illustration of vLLM's Paged Attention memory blocks and the preemption mechanism when GPU memory becomes insufficient; the newest request is preempted, its KV-cache blocks are released, and it is later recomputed and resumed.}
    \label{fig:pageAttention}%
\end{figure}

A key challenge is memory allocation: response lengths are unknown at runtime, so pre-allocating for max sequence length fragments memory and wastes it on short responses. Paged Attention~\cite{vllm}, implemented in vLLM, partitions GPU memory into fixed-size blocks and tracks each KV cache via a page table, allowing non-contiguous placement (Figure~\ref{fig:pageAttention}). When memory is insufficient for the next decode step, vLLM preempts the newest in-batch request, releases its blocks, and later resumes it by recomputing the KV cache---a key reason for vLLM's practical effectiveness.

\begin{figure}
    \centering
    \includegraphics[width=1.0\linewidth]{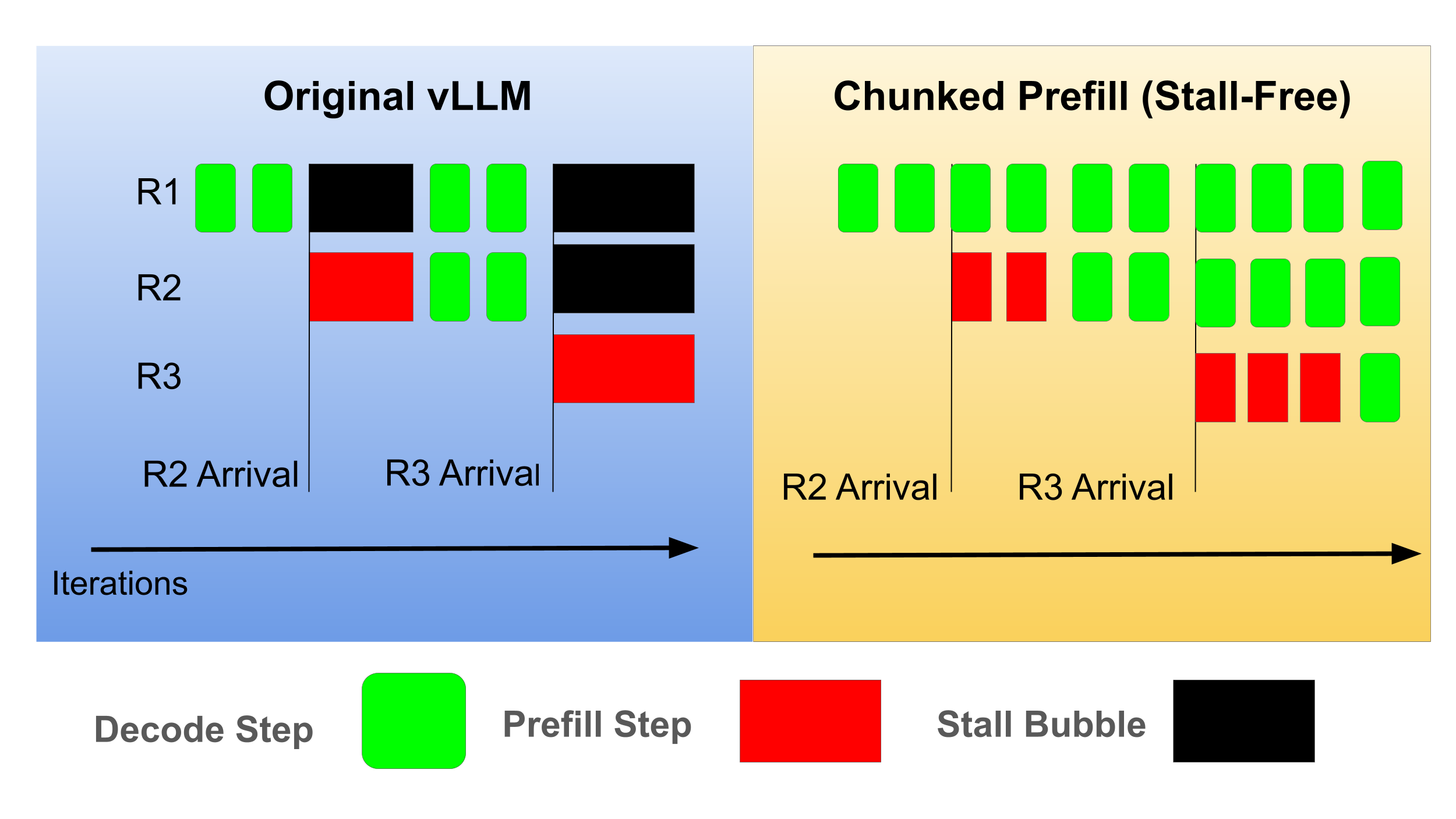}
    \caption{Original vLLM and Chunked Prefill}
    \Description{Comparison of vLLM's original prioritized prefill scheduling with chunked prefill, highlighting how chunked prefill interleaves prefill chunks with decode to reduce decode stalls and tail latency.}
    \label{fig:chunked_prefill}
\end{figure}

Hybrid prefill and decode scheduling is the next challenge. vLLM's original scheduler creates separate prefill-only and decode batches and prioritizes prefill (higher throughput, lower TTFT), but this interrupts decode batches and causes decoding stall bubbles and tail-latency regressions (Figure~\ref{fig:chunked_prefill}).

Chunked Prefill with a stall-free local scheduler~\cite{agrawal2023sarathiefficientllminference, agrawal2024tamingthroughputlatencytradeoffllm} mitigates this trade-off: prompt processing is split into equal-sized chunks executed across scheduling steps, forming hybrid batches that interleave decode steps with prefill chunks under a token budget, configured as the chunk size. This reduces decode stalls with only minor throughput loss, making it the default in vLLM and SGLang~\cite{sglang}. Prefill-Decode (P-D) disaggregation~\cite{disServe, patel2024splitwiseefficientgenerativellm, qin2024mooncakekvcachecentricdisaggregatedarchitecture} is an alternative that uses separate instances at the cost of cross-instance KV-cache transfers. However, aggregated execution can still be preferable in some settings~\cite{wang2025prefilldecodeaggregationdisaggregationunifying}, and GPU multiplexing~\cite{gao2025duetserveharmonizingprefilldecode} offers similar isolation benefits without disaggregation's KV-cache transfer overhead.

\section{Related Work and Motivation}\label{section:scheduling_issues}

Current LLM inference with dynamic batching and memory allocation, as discussed above, introduces new cloud-based challenges for task scheduling and load balancing. Accordingly, this section first elaborates on these issues and then summarizes applicable techniques, leading to the proposed solution discussed in \S\ref{section:system_design}.

\begin{figure}
    \centering
    \includegraphics[width=1.0\linewidth]{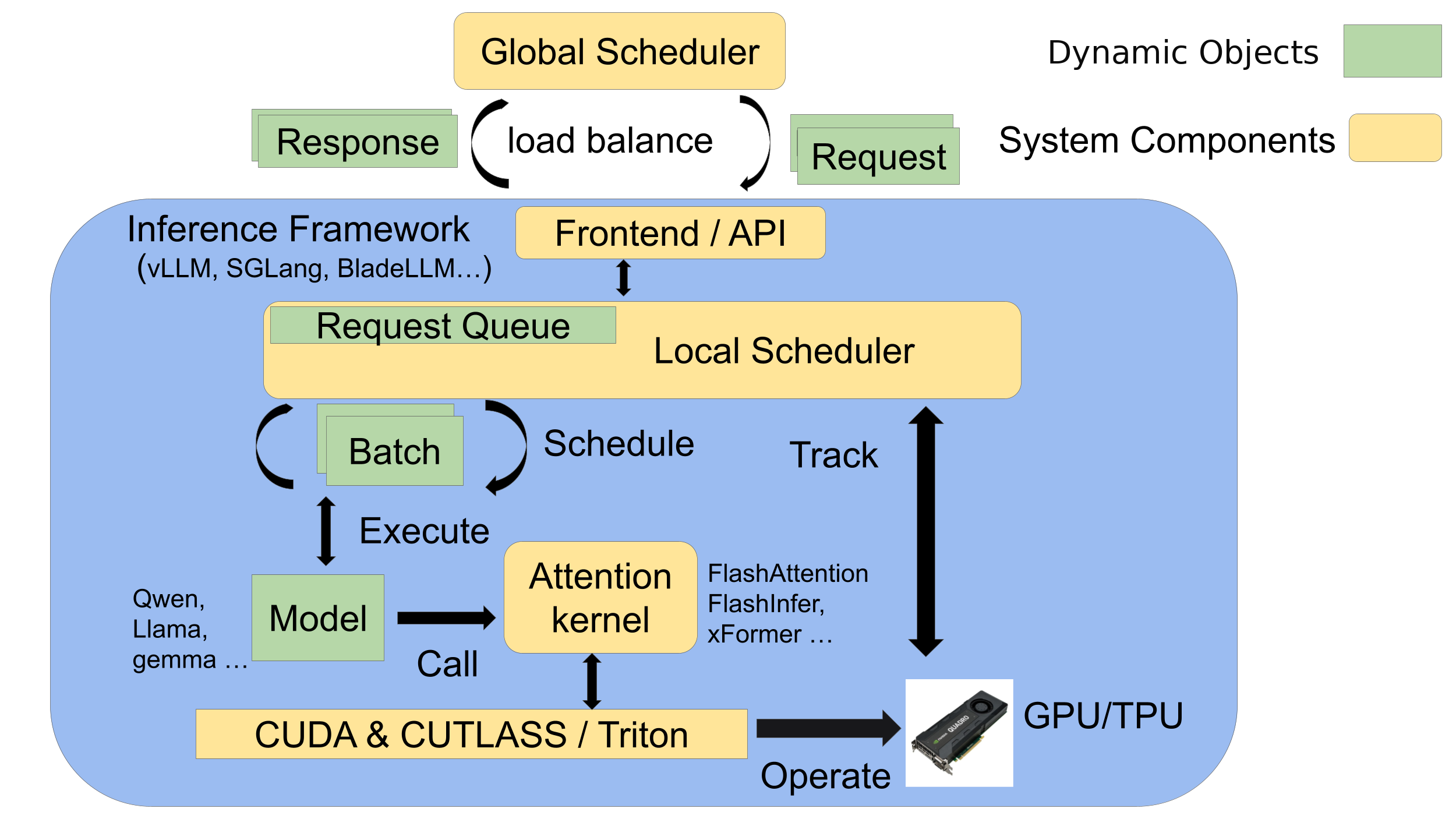}
    \caption{LLM Inference Systems}
    \Description{High-level workflow of an LLM inference serving system, including the user/API frontend, local batching scheduler, model execution, and a global scheduler routing requests across multiple instances.}
    \label{fig:llm-inference-workflow}
\end{figure}

\paragraph{Unpredictability in LLM Serving Scheduling} \label{section:unpredictability} A typical architecture for LLM inference serving can be seen in Figure~\ref{fig:llm-inference-workflow}.  First, a user interacts with an API that serves as the entry point to the inference framework. A local scheduler within the framework then batches these incoming requests for parallel execution. The LLM inference process is composed of a series of layers, primarily involving attention mechanisms and linear projections. To optimize performance, these operations are often accelerated by highly efficient kernel implementations~\cite{flashattentionv2, ye2025flashinfer}. Real-world services~\cite{wang_gemini_2023,claude, achiam2023gpt} typically deploy multiple framework instances for availability under heavy load, requiring an external global scheduler to balance requests across instances.

LLM inference systems are typically characterized by high unpredictability in both the final outputs and overall system performance over time, as noted in~\cite{sun2024llumnix}. This unpredictability stems primarily from two sources: 1) variable memory demands caused by unknown decoding lengths, which can trigger unexpected preemption; and 2) resource competition, which can create interference between colocated requests. Both effects can degrade runtime performance.

This unpredictability makes dispatching across LLM instances more challenging than general dynamic task scheduling~\cite{dynamic_task_scheduling}: a scheduler like Kubernetes filters nodes on available resources and picks the least utilized, but such static filtering and utilization metrics are inadequate when request memory demands vary dynamically with decoding length. Additionally, the actual performance when executing a request can still be affected by interference from other colocated requests~\cite{sun2024llumnix}. For example, unexpected preemption under GPU-memory contention can reset active requests and block later ones. In contrast, traditional cloud tasks run in isolated, statically allocated containers~\cite{docker}, which limits interference and lets task lifetimes be predicted from history, as in Apollo~\cite{Apollo} and LAVA~\cite{ling2025lavalifetimeawarevmallocation}.

These factors make it difficult to accurately predict resource demands and duration of a request on a given instance. Consequently, many dispatching strategies across LLM instances rely on simple heuristics such as round-robin (DeepSpeed-MII~\cite{deepspeed}, Triton Inference Server~\cite{nvidiaTritonInference}), while vLLM delegates cross-instance routing to the user or a higher-level orchestrator. SGLang ships its own heuristic router coupled with internal optimizations such as RadixAttention; ServerlessLLM~\cite{serverlessLLM} schedules requests to instances with the least estimated model-loading time; and LLM-d~\cite{githubGitHubLlmdllmd} serves LLMs on Kubernetes with separate prefill/decode schedulers for P-D disaggregation. 

These rule-based heuristic schedulers can be tuned to perform well under specific configurations, workloads, or models, but they could degrade in dynamic environments and lack quantifiable metrics linking internal rules to user-facing performance. In general, workload unpredictability and the absence of accurate dispatch-time metrics make it hard for heuristics to balance load consistently, especially under high, variable load. Llumnix~\cite{sun2024llumnix} proposes a dynamic re-balancing approach to mitigate such load imbalance by migrating requests across instances with their KV caches. However, migration can introduce significant overhead and network contention, and can degrade performance and even collapse the entire cluster under high load. This motivates the need for more proactive one-shot scheduling approaches that can improve load balancing without migration.

\paragraph{Offline Performance Simulator} \label{section:performance simulation}

With the rapid development and expansion of new LLM models, devices, and use cases, setting up an appropriate LLM serving cluster has become a critical yet challenging task for developers, particularly because exploring candidate configurations by trial and error directly on GPUs is expensive. To address this, Vidur, an LLM serving cluster simulation framework, has been proposed~\cite{vidur}. It aims to reduce potential hardware costs when searching for the optimal cluster configuration based on a given model, trace, and service-level objective (SLO) requirements. The insight behind Vidur is that since the local scheduler and batching logic are deterministic, and response lengths are typically available in the replay trace, it is feasible to simulate the entire replay process if the execution time for each batch can be inferred. Additionally, by profiling low-level operators like attention and linear projection for a specific GPU, Vidur trains linear models to interpolate execution times for various batches. It reports less than 9\% error for key metrics such as throughput and latencies. Other frameworks, such as LLMServingSim~\cite{cho2025llmservingsim}, Frontier~\cite{feng2025frontier}, Revati~\cite{agrawal2026revati}, and LLM-Emu~\cite{da2026llmemunativeruntimeemulation} have also been proposed to simulate LLM serving clusters with different areas of focus.

\paragraph{Response Length Prediction} \label{section:length_prediction} Prior work predicts unknown response lengths to help \emph{local} schedulers: Sequence Scheduling~\cite{length_estimation} fine-tunes another LLM to predict response length from the input instruction and batches similar-length requests to achieve 85\% higher throughput. LightLLM~\cite{lightLLM} uses historical distributions to estimate the expected length to avoid preemption. The ELIS~\cite{ELIS} local scheduler uses a regression model to estimate the length to construct batches with shorter completion times. Recent work applies predictive scheduling to inference-time reasoning, estimating per-query reasoning budgets~\cite{brown2026predictivescheduling}.

Length estimation has also been utilized for global scheduling, primarily as a filtering mechanism prior to scheduling tasks. For instance, TetriServe~\cite{TetriServe} deploys a lightweight length-estimation model to filter out decoding candidate instances without sufficient GPU memory to accommodate the estimated tokens and then applies the power-of-two method~\cite{poweroftwo} to select decoding instances with fewer pending requests. DynamoLLM~\cite{dynamollm} classifies requests into three distinct pools based on their estimated lengths, scheduling them separately with associated host pools. To summarize, our insight is that by combining simulation techniques with length prediction, the unpredictability can be substantially reduced, which in turn creates an opportunity to improve multi-instance LLM serving performance.

\begin{figure}[!tb]
    \centering
    \includegraphics[width=1.0\linewidth]{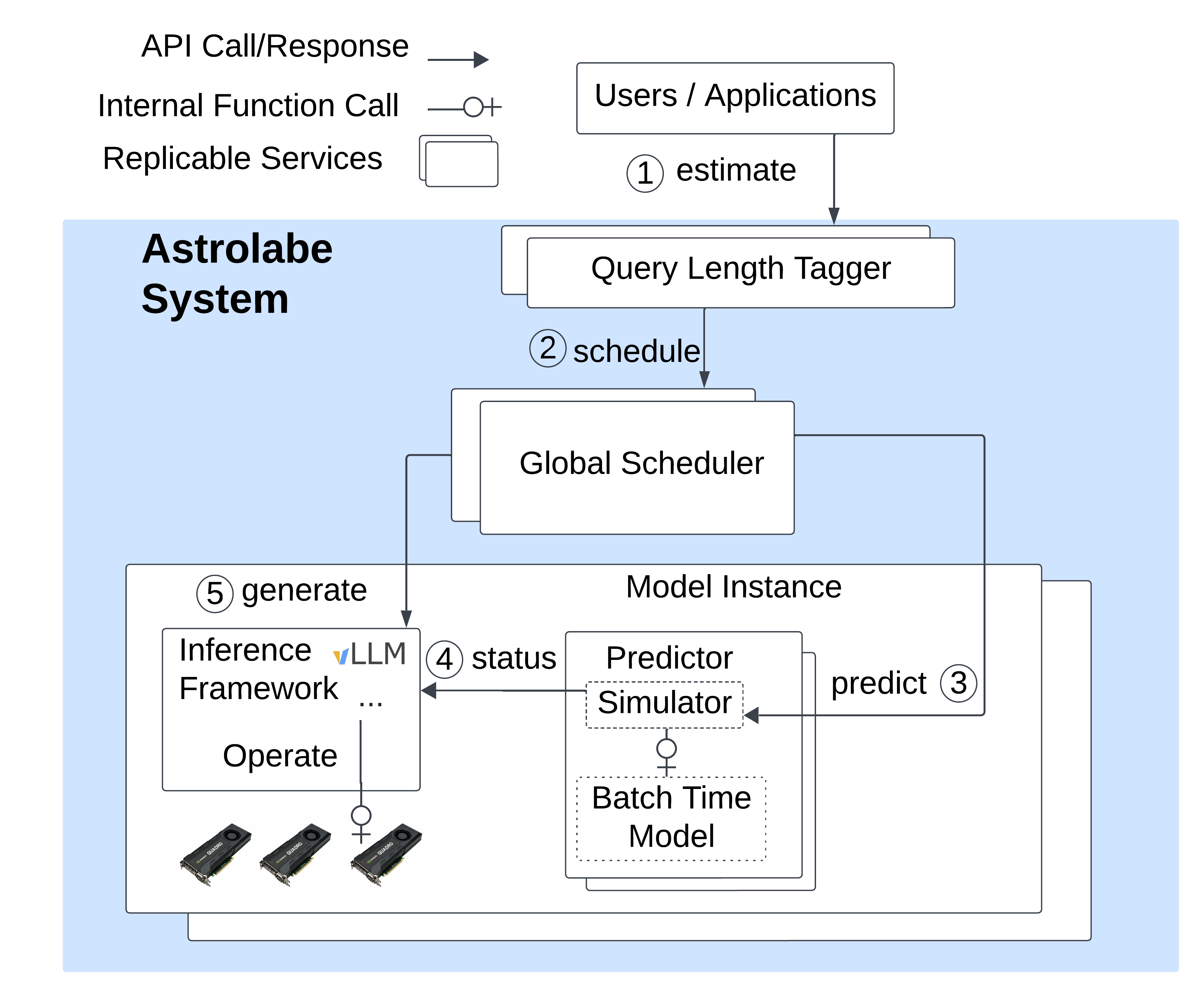}
    \caption{Astrolabe Architecture}
    \Description{System architecture of Astrolabe, showing four components (query length tagger, global scheduler, predictor sidecars attached to each model instance, and the inference framework backend) and the flow from incoming requests through prediction, scheduling, execution, and returning results.}
    \label{fig:scheduling_framework}
\end{figure}

\section{Astrolabe System Design} \label{section:system_design}

\emph{Astrolabe} is a prediction-guided task scheduler that leverages query context and cluster knowledge for load balancing in LLM serving clusters. Building on the insight from \S\ref{section:scheduling_issues}, it couples runtime prediction with a randomized dispatch policy for high-quality load balancing.

Astrolabe, as shown in Figure~\ref{fig:scheduling_framework}, comprises four services: a query length tagger, a global scheduler, per-instance Predictor sidecars, and inference framework backends. The query length tagger is the entry point and predicts the anticipated response length for each request (\Circled{1}). Requests are then forwarded to the global scheduler (\Circled{2}), which selects a target model instance.
For each incoming request, the global scheduler issues \texttt{predict} RPCs to a sampled subset of Predictor sidecars (\Circled{3}) (power-of-$k$ choices; $k{=}2$ by default, and $k{=}N$ recovers full fanout). Each Predictor queries its colocated backend via a \texttt{status} API to obtain the instance's current runtime state (e.g., running/waiting requests, batching state, and memory availability), and then runs a lightweight simulation to estimate the request's target metric (e.g., E2E latency) (\Circled{4}). The global scheduler picks the target instance from these predictions per the configured policy (\Circled{5}), and the backend executes the request and returns the response. Astrolabe is designed as an add-on scheduling layer and is agnostic to models, hardware, inference frameworks, and dispatch policies.

\subsection{Model Instance} \label{section:model_instance}

A model instance is the collection of services on a GPU host driving one or more GPUs: the inference framework itself plus a sidecar service called the Predictor. Astrolabe is framework-agnostic and supports dynamic-batching frameworks~\cite{vllm, orca, deepspeed} already covered by Vidur.
Each inference framework used by Astrolabe can be integrated by supporting a new \texttt{status} API to export its internal status, such as request lists and GPU memory blocks, for metric prediction.

The Predictor's main role is to predict key performance metrics, such as E2E latency or TTFT, for incoming requests.
The Predictor runs locally on each instance, aggregates runtime data from the \texttt{status} API, and serves a scalar metric prediction via the \texttt{predict} API to the global scheduler.
In cases where a request's decoded length exceeds its predicted length at runtime, the simulator extends the current prediction by assuming the request will complete within the next 10 tokens by default.

Astrolabe also works with different simulation approaches. In our prototype we integrate Vidur~\cite{vidur}, redesigned for single-instance prediction and encapsulated behind the Predictor API. As illustrated in Figure~\ref{fig:scheduling_framework}, simulation proceeds in two stages: a local-scheduler simulator models the backend's batching strategy, then a linear model predicts execution times for the generated batches. Since the model is static and inputs are received on demand from the \texttt{status} API, the Predictor service does not maintain scheduling state and can be replicated per instance to reduce simulation-related overhead through parallelization. Furthermore, the Predictor service is designed with an extensible interface, which could be implemented by alternative simulation frameworks while keeping core strategies.

\subsection{Global Scheduler} \label{section:schedulers}

\begin{algorithm}[t]
\caption{Astrolabe one-shot dispatch (default: latency)}
\label{alg:astrolabe_dispatch}
\begin{algorithmic}[1]
\Require Request $r$ with prompt $p$; instances $\mathcal{I}=\{i_1,\dots,i_N\}$; sample size $k$ (default $2$); timeout $\tau$
\Ensure Dispatched instance $i^\star$
\State $\hat{L} \gets \textsc{LengthTagger}(p)$
\State $\mathcal{I}' \gets \textsc{Sample}(\mathcal{I}, k)$
\State $\mathsf{cand} \gets \emptyset$
\ForAll{$i \in \mathcal{I}'$ \textbf{in parallel}}
    \State $s_i \gets \textsc{GetStatus}(i)$
    \State $\hat{\ell}_i \gets \textsc{PredictMetric}(i, r, \hat{L}, s_i)$
    \If{$\hat{\ell}_i$ returned before $\tau$}
        \State $\mathsf{cand} \gets \mathsf{cand} \cup \{(i,\hat{\ell}_i)\}$
    \EndIf
\EndFor
\If{$\mathsf{cand}=\emptyset$}
    \State \textbf{reject} $r$ and \textbf{throw exception}
\EndIf
\State $i^\star \gets \arg\min_{(i,\hat{\ell}_i)\in \mathsf{cand}} \hat{\ell}_i$
\State \textsc{Dispatch}$(i^\star, r)$
\end{algorithmic}
\end{algorithm}

The global scheduler service is stateless and lightweight without maintaining a global table of instance statuses. Instead, as illustrated in Figure~\ref{fig:scheduling_framework}, upon each request arrival, the scheduler issues a \texttt{predict} RPC to sampled Predictor sidecars and then dispatches the request to the instance with the lowest predicted metric. In this paper, we use E2E latency as the default target metric, while keeping the interface configurable to support other objectives if needed.

This differs from centralized predictors that need a global per-instance status view (e.g., the estimation-driven global schedulers in TetriServe~\cite{TetriServe} and DynamoLLM~\cite{dynamollm}). In Astrolabe, prediction runs on each instance in parallel using its local state, and each Predictor returns only a scalar metric (e.g., predicted latency). Decision time is bounded by the slowest Predictor reply, while avoiding the cost of exporting heavy instance state (e.g., request lists) to a central simulator. Because each Predictor simulates the request on top of its execution queue, the prediction reflects both running and pending load.

To avoid the $O(N)$ per-dispatch messaging cost of full fanout and the scheduler-herding issue (\S\ref{section:introduction}), Astrolabe's default samples $k{=}2$ random instances (power-of-two choices) and queries only those Predictors. Setting $k{=}N$ recovers full fanout; we sweep $k$ in \S\ref{section:po2_ablation}. The dispatch procedure is shown in Algorithm~\ref{alg:astrolabe_dispatch}.

Furthermore, Astrolabe does not require instance-side features such as live migration for dynamic rebalancing~\cite{sun2024llumnix}. Such features can be complementary in some deployments, but they introduce additional orchestration and state-transfer overhead. In this work, we focus on prediction-guided one-shot dispatch and show that it can match or outperform migration-based rebalancing.

\subsection{Query Length Tagger}\label{section:query_tagger}
The query length tagger is an online service that estimates the response length of each request from its prompt and the target serving model. It runs in parallel with request handling and uses a lightweight proxy model to keep overhead low; the estimator is modular and can be replaced with alternatives such as the model-free, sampling-based approach used in LightLLM~\cite{lightLLM}. The predicted length is used only for metric prediction and scheduling decisions and does not change the inference parameters and results.

\section{Implementation} \label{section:implementation}

Astrolabe implements the global scheduler, per-instance Predictor sidecars, and the length-tagger service (\S\ref{section:system_design}) as lightweight FastAPI~\cite{tiangoloServerWorkers} microservices with a simple RPC/REST interface that integrates with common LLM serving stacks (e.g., vLLM~\cite{vllm}, SGLang~\cite{sglang}). The prototype integrates with vLLM 0.7.2 and comprises about 4{,}000 LoC, released at \url{https://github.com/AKafakA/Block}.

\paragraph{Simulation-based prediction}
The Predictor is the per-instance sidecar described in \S\ref{section:model_instance}; internally it invokes a backend-specific simulator. Our prototype reuses Vidur's offline simulator codebase~\cite{vidur} as the starting point, due to its high-fidelity modeling, broad backend support, and extensibility (Sarathi-Serve, vLLM, LightLLM), and substantially adapts it for online, per-request use. Vidur's offline pipeline has inefficiencies that are negligible offline but become critical for real-time prediction (e.g., object duplication, suboptimal \texttt{list.pop(0)} calls), so we rewrote the primary simulation hot paths and added a per-Predictor cache that memoizes latency predictions per (batch size, token count), substantially reducing the per-request simulation cost.

\paragraph{Framework Integration and Global Scheduler Implementation} \label{section:baselinescheduler}
Astrolabe is decoupled from any specific inference framework (\S\ref{section:model_instance}): integrating a new backend requires only exposing the read-only \texttt{status} API and implementing a local-scheduler simulator. Extending to new serving features (e.g., P-D disaggregation) similarly requires updating only the simulator and, if needed, the state exported via \texttt{status}.

Astrolabe's global scheduler exposes a configurable target metric and dispatch policy, supporting single- and multi-SLO strategies~\cite{chen2025slosserveoptimizedservingmultislo}. Our prototype selects the instance with the lowest predicted E2E latency. For evaluation we implement five baselines within the same framework:
\begin{itemize}
    \item \emph{Random}: selects an instance uniformly at random.
    \item \emph{Round-Robin}: cyclic dispatch, used by production-grade serving systems~\cite{nvidiaTritonInference, rayServeScalable}.
    \item \emph{Min QPM (Queries Per Minute)}: LiteLLM's~\cite{litellm} default policy to dispatch based on QPM as a proxy for load.
    \item \emph{INFaaS++}: the Llumnix-optimized variant of INFaaS~\cite{sun2024llumnix,infaas}, with load as $usedMemory/batchSize$.
    \item \emph{Llumnix-} (the trailing dash marks the scheduler-only variant): Llumnix's~\cite{sun2024llumnix} heuristic dispatcher extending INFaaS++ with prefill memory correction as:
    
    $load = (usedMemory + prefillMemory) / batchSize$.
    Live migration is disabled in the main evaluation and full migration-enabled Llumnix is evaluated in~\S\ref{section:llumnix_comparison}.
\end{itemize}

\paragraph{Length Estimation Model}
For the query length tagger (\S\ref{section:query_tagger}), we fine-tuned a lightweight RoBERTa-base~\cite{roberta} regression model (125M parameters); similar BERT-based estimators are widely adopted in serving systems for auxiliary tasks ranging from model routing~\cite{wang2025reasonsemanticroutervllm, ong2025routellmlearningroutellms, ding2025bestrouteadaptivellmrouting} to output-length prediction~\cite{ELIS, s3}.~\footnote{We initially considered the 7B length model from Sequence Scheduling~\cite{length_estimation}, but its training duration and serving overhead were prohibitive.}

\paragraph{Use of AI Tools}
Astrolabe was designed and implemented by the authors, and both the system prototype and the initial version of this paper predate the project's adoption of generative-AI coding tools. In later stages we used AI assistance in a supporting role, which we disclose per the ACM authorship policy: in-editor code-completion suggestions during implementation iterations, and AI coding agents that helped write experiment-automation and plotting scripts and orchestrate and monitor long-running experiments on the remote clusters; AI tools were also used to polish the paper's prose. All AI-assisted code and text were reviewed, tested, and validated by the authors, who take full responsibility for them; the system design, experimental methodology, data analysis, and all conclusions are the authors' own.

\section{Evaluation} \label{section:evaluation}
We evaluate Astrolabe against five baseline schedulers (\S\ref{section:baselinescheduler}). We introduce the experimental setup at \S\ref{section:experiment_setting}, validate prediction accuracy at \S\ref{section:predicting accuracy}, and quantify E2E gains in \S\ref{section:request_performance} (including GPU memory dynamics and a head-to-head comparison against Llumnix at \S\ref{section:llumnix_comparison}). We then present additional analyses on power-of-$k$ choices, with $k{=}2$ by default (\S\ref{section:po2_ablation}), oracle-length upper bounds (\S\ref{section:oracle_bounds}), burstiness (\S\ref{section:burstiness}), prediction-noise sensitivity (\S\ref{section:error_sensitivity}), and CPU overhead (\S\ref{section:cpu_overhead}).

\subsection{Experimental Setup} \label{section:experiment_setting}

\paragraph{Testbed} We evaluate Astrolabe on CloudLab~\cite{cloudlab} using two clusters. The primary cluster consists of 12 d7525 nodes, each equipped with two 16-core AMD 7302 CPUs (3.00\,GHz), 128\,GB ECC memory, one NVIDIA A30 GPU (24\,GB), and 25\,Gbps NICs (best-effort CloudLab provisioning, so nominal 25\,Gbps is an upper bound). This single-GPU-per-host configuration is used for the majority of experiments using the implementation and baselines described in \S\ref{section:implementation}.

Additionally, we deploy a 2-node CloudLab d8545 cluster with 4$\times$ A100-40GB GPUs per node and 100\,Gbps NICs. This 4-GPU-per-node configuration is comparable to Llumnix's original testbed~\cite{sun2024llumnix}. The multi-GPU cluster enables: (1) comparison with full Llumnix migration (\S\ref{section:llumnix_comparison}) with KV-cache transfers; and (2) evaluation of prediction accuracy on larger models with tensor parallelism (\S\ref{section:predicting accuracy}). We run the length-estimation model on a dedicated host with an L40 GPU to process datasets offline. Such lightweight BERT-based models can run online, as in recent production LLM services~\cite{wang2025reasonsemanticroutervllm, ong2025routellmlearningroutellms}; we run them offline to avoid interference with the evaluation and to iterate faster. The RoBERTa-base tagger (125M parameters) is fine-tuned once per model--tokenizer pair on 40k ShareGPT pairs as a one-time offline cost on a single GPU.

\paragraph{Data, Configuration and Model}
We use ShareGPT~\cite{shareGPTData}, a real-world dataset with 50K conversations for the main experiments. During experiments, prompts are sent to the global scheduler according to a Poisson arrival process at varying rates, referred to below as external QPS.

When serving LLMs on vLLM, we use FlashInfer~\cite{ye2025flashinfer} as the attention library and greedy decoding~\cite{jm3} with temperature 0. The maximum request batch size is 48, and the Chunked-Prefill chunk size is 512; this combination gives the best performance among tested configurations. To reduce simulation overhead, each instance's node runs 16 Predictor replicas to simulate and predict latencies in parallel, which reduces total scheduling overhead by up to 50\%.

Since each A30 test node has a single GPU and limited cross-node bandwidth, co-serving across GPUs using tensor or pipeline parallelism~\cite{narayanan2021efficientlargescalelanguagemodel} is infeasible. We therefore use a medium-sized model that fits within a single node: Llama-2-7B~\cite{llama} with FP16 precision. We choose Llama-2 to stay aligned with Vidur~\cite{vidur}, which lets us reuse Vidur's published model configuration when profiling operators for our simulator; the Llama family is also widely used in prior AI/systems papers. The model weights occupy 12.5\,GB of GPU memory, split into 1056 vLLM memory blocks for KV cache. For tensor-parallel prediction accuracy, we also evaluate Llama-2-70B on the A100 cluster (\S\ref{section:predicting accuracy}) and observe comparable latency-prediction accuracy (MAPE 16.6\%), suggesting that Astrolabe's predictor generalizes to production-scale deployments. Vidur~\cite{vidur} also reports that simulation accuracy can improve with larger models because framework CPU overhead accounts for a smaller fraction of E2E latency.

\subsection{Prediction Accuracy} \label{section:predicting accuracy}

\begin{table}[tb]
    \caption{Response Length Prediction. Acc-X: Ratio of token predictions with absolute error < X.}
    \centering
    \begin{tabular}{ccc}
    \toprule
    Metric & RoBERTa Regressor & Prompt-based LLM  \\ \midrule
    MAE & 78.755 & 62 \\
    MAPE & 24.4\% & Not reported \\
    Acc-50 & 69.93\% & 59\% \\
    Acc-100 & 77.15\% & 81\% \\ 
    \bottomrule
    \end{tabular}    \label{table:length_prediction}
\end{table}%

\paragraph{Length Prediction.}
We divide the first 50k ShareGPT records into 40k training and 10k evaluation samples. The labels are actual output lengths generated by the served model, and we use them to fine-tune a RoBERTa-base regression model. The resulting 125M-parameter model reaches MAE 78.8 and MAPE 24.4\% on the 10k test split, which is comparable to the 7B prompt-based model reported in Sequence Scheduling~\cite{length_estimation} (Table~\ref{table:length_prediction}; updated numbers from~\cite{githubGitHubZhengzangwSequenceScheduling}). Offline evaluation on 10k requests completes in 4.8\,s in PyTorch, consistent with online overheads reported for comparable BERT-based estimators in other runtime systems ($S^3$: 3.7\,ms~\cite{s3}; ELIS: 11.4\,ms incl.\ batching~\cite{ELIS}).

\paragraph{Simulation-based Metric Prediction.}
\begin{figure*}[!tb]
    \centering
    \includegraphics[width=0.93\linewidth]{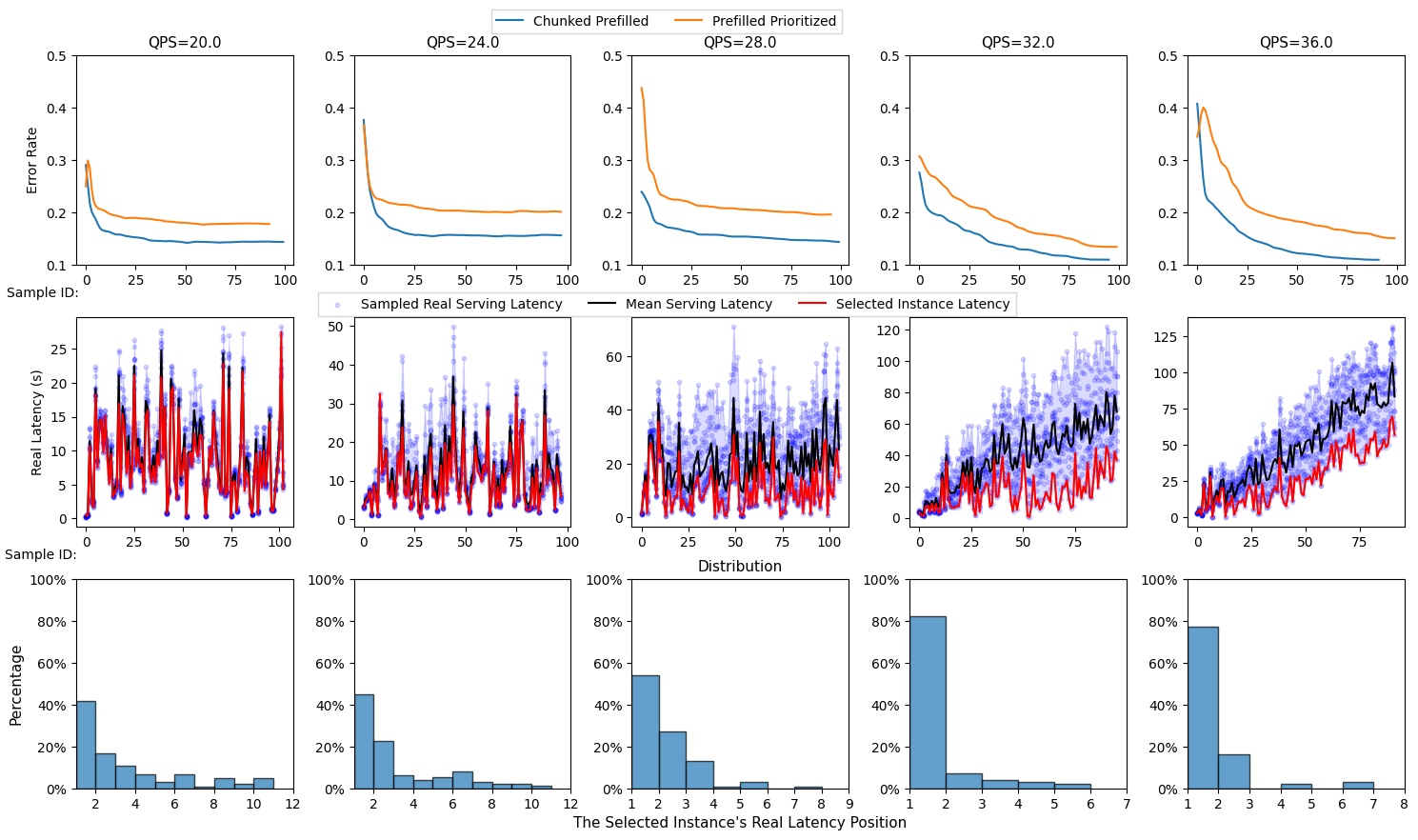}
    \caption{Latency Prediction Metrics}
    \Description{Metrics summarizing latency prediction accuracy and scheduling effectiveness, including prediction error and the rank of the instance selected by the scheduler among all candidates.}
    \label{fig:latency_prediction}
\end{figure*}

Figure~\ref{fig:latency_prediction} assesses the runtime simulator's accuracy and its effectiveness for scheduling. Astrolabe processes requests with real lengths at QPS values from 20 to 36 using a random scheduler, recording predicted and actual latencies. Online requests are sampled at 1\%. For each sampled request, we broadcast the request to all instances to collect predicted and actual latencies, then mark the minimal-predicted-latency instance as selected.

The top row in the figure reports the prediction error rate. Enabling Chunked Prefill yields much lower error than vLLM's default Prioritized Prefill, whose decoding stalls are triggered by future arrivals that are unknown at prediction time and hence unmodelable. Chunked Prefill mitigates this by interleaving prefill and decode phases as explained in \S\ref{section:motivation}, aligning runtime behavior with the simulator's linear model. The middle and bottom rows further validate the scheduling effectiveness. The scatter plots in the middle row show that the latency of the selected instance (red line) consistently tracks the lower envelope of the latency distribution (blue dots) and remains significantly below the cluster mean (black line). Even when not selecting the absolute best instance, the scheduler consistently avoids overloaded nodes. The bottom row displays the 'Real Latency Position', denoting the rank of the selected instance among all candidates (where 1 is optimal). While the absolute prediction error ranges from 10 to 15\%, Astrolabe mainly relies on \emph{relative} latency ranking for instance selection: the scheduler maintains a high probability (40 to 80\%) of selecting the Rank-1 instance. This is because a large fraction of prediction error manifests as near-constant system noise (e.g., RPC overhead) that affects all instances similarly and therefore does not change their relative ordering.

To validate that Astrolabe's simulation accuracy extends to larger models with tensor parallelism, we evaluate Llama-2-70B on the 2-node A100 cluster with TP=4 (2 instances, 4 GPUs each). At QPS=3 (scaled to this two-instance TP=4 deployment) with a 10\% sampling rate, the simulator achieves a latency-prediction error of 16.6\%. This 16.6\% is well within Astrolabe's robustness envelope: \S\ref{section:error_sensitivity} shows the scheduler tolerates up to $\pm$100\% independent Gaussian noise on simulator latency with at most +6.3\% mean E2E degradation and +3.7\% P99 E2E degradation across the heatmap, because Po2 dispatches on \emph{relative} ranking and common-mode error cancels as discussed later. Vidur~\cite{vidur} additionally reports that simulator accuracy may improve at larger model scales because unmodeled per-request CPU overhead becomes a smaller fraction of E2E latency.

\begin{figure*}[!tb]
        \centering
          \includegraphics[width=0.93\linewidth]{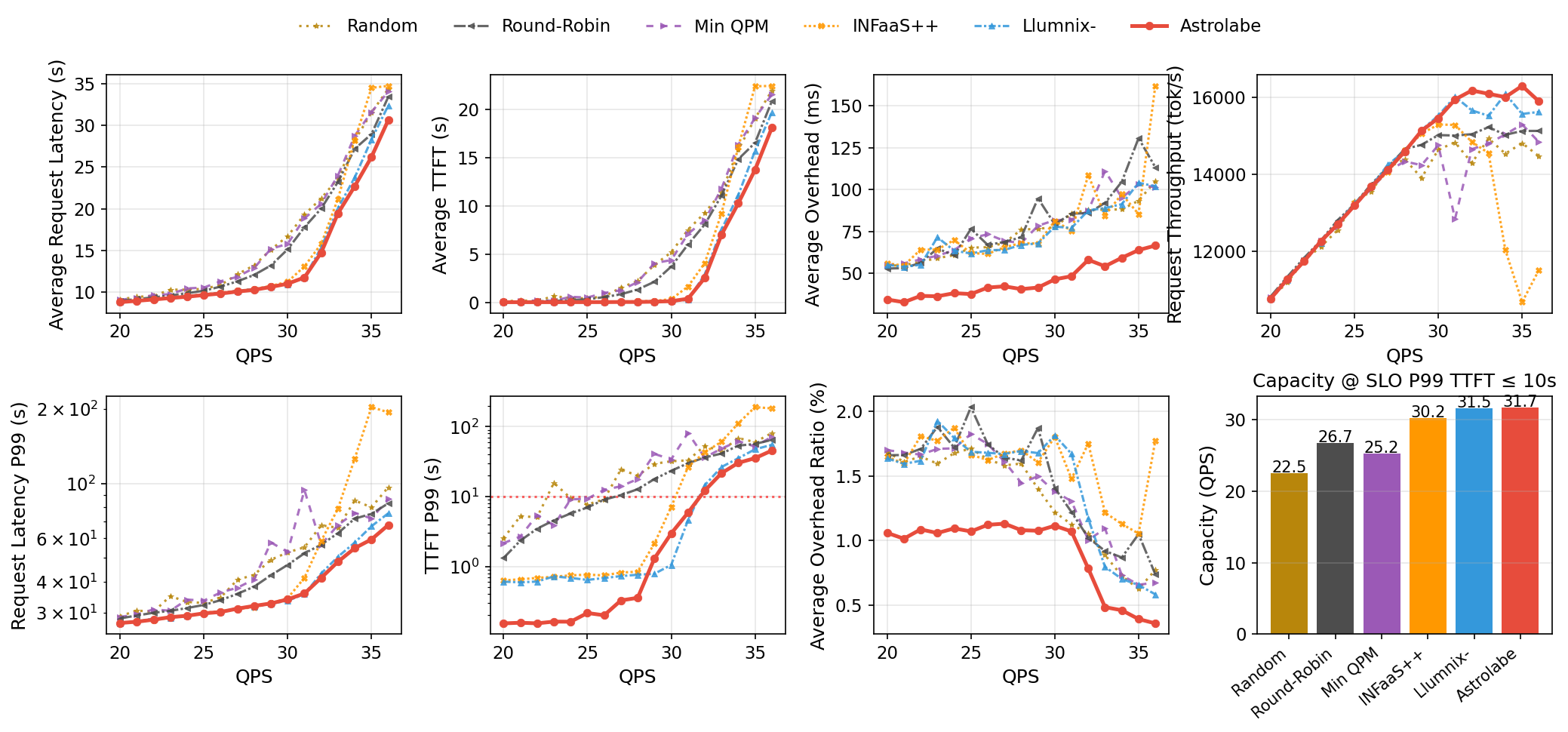}
        \caption{Request Metrics Under Different QPS}
        \Description{Key serving metrics (e.g., E2E latency, TTFT, tail latency, throughput, and scheduling overhead) measured across different external query-per-second (QPS) levels for Astrolabe and baseline schedulers.}
        \label{fig:request_metrics}
\end{figure*}

\subsection{Request Latency Performance} \label{section:request_performance}
Figure~\ref{fig:request_metrics} evaluates Astrolabe ($k{=}2$, RoBERTa lengths) against five non-predictive baselines as discussed in \S\ref{section:baselinescheduler} across an integer-QPS sweep. We report E2E latency, TTFT, scheduling overhead, serving \emph{capacity} (max QPS meeting SLO P99 TTFT\,$\le$\,\sloSec\,s, 0.1-QPS resolution), and request throughput. Llumnix- is the primary non-predictive comparator. Table~\ref{tab:request_metrics_snapshot} (snapshot at QPS=20, 24, 28, 32, 36) quantifies the latency metrics in more detail and reports Astrolabe's gain over the best non-Astrolabe scheduler at each QPS. Each (scheduler, QPS) point is a single 10k-request run. To bound deployment variability we repeated this sweep on separately provisioned clusters: across all six schedulers, serving capacity reproduces within 0.6\,QPS, and mean throughput and E2E latency within 2.4\% and 1.7\%. Differences of a few percent in Table~\ref{tab:request_metrics_snapshot} are therefore within deployment noise, while the ordering across schedulers is stable; tail TTFT is noisier.

\begin{table*}[!tb]
\caption{Per-scheduler TTFT (ms)/E2E (s); last row: Astrolabe's gain over the best non-Astrolabe scheduler.}
\centering
\begingroup
\setlength{\tabcolsep}{3.6pt}
\renewcommand{\arraystretch}{1.02}
\resizebox{\textwidth}{!}{%
\begin{tabular}{l ccccc c ccccc c ccccc c ccccc}
\toprule
& \multicolumn{5}{c}{TTFT mean (ms)} & & \multicolumn{5}{c}{TTFT P99 (ms)} & & \multicolumn{5}{c}{E2E mean (s)} & & \multicolumn{5}{c}{E2E P99 (s)} \\
\cmidrule(lr){2-6} \cmidrule(lr){8-12} \cmidrule(lr){14-18} \cmidrule(lr){20-24}
QPS $\to$ & 20 & 24 & 28 & 32 & 36 & & 20 & 24 & 28 & 32 & 36 & & 20 & 24 & 28 & 32 & 36 & & 20 & 24 & 28 & 32 & 36 \\
\midrule
Astrolabe & \textbf{67} & \textbf{71} & \textbf{88} & \textbf{2586} & \textbf{18107} & & \textbf{153} & \textbf{161} & \textbf{355} & \textbf{12152} & \textbf{45506} & & \textbf{8.8} & \textbf{9.4} & \textbf{10.3} & \textbf{14.7} & \textbf{30.6} & & 27.3 & \textbf{29.2} & 32.0 & \textbf{41.6} & \textbf{67.9} \\
Llumnix-  & 96 & 111 & 133 & 2935 & 19707 & & 605 & 684 & 759 & 14505 & 54989 & & 8.8 & 9.5 & 10.4 & 15.3 & 32.4 & & \textbf{27.2} & \textbf{29.2} & 31.8 & 43.3 & 75.8 \\
INFaaS++  & 106 & 131 & 149 & 4111 & 22456 & & 630 & 745 & 840 & 43176 & 182614 & & 8.8 & 9.5 & 10.4 & 15.8 & 34.7 & & \textbf{27.2} & \textbf{29.2} & \textbf{31.6} & 58.6 & 193.9 \\
RR        & 99 & 279 & 1381 & 8113 & 20857 & & 1320 & 5719 & 12735 & 36266 & 64683 & & 9.0 & 9.9 & 12.1 & 20.0 & 33.4 & & 28.5 & 31.4 & 38.3 & 56.6 & 83.1 \\
MinQPM    & 127 & 624 & 2153 & 8538 & 21559 & & 2154 & 9118 & 17788 & 36160 & 71727 & & 9.1 & 10.4 & 12.9 & 20.6 & 34.0 & & 28.9 & 33.9 & 41.0 & 56.6 & 86.6 \\
Random    & 158 & 501 & 2279 & 9309 & 21969 & & 2529 & 9485 & 19881 & 53212 & 81112 & & 9.1 & 10.3 & 13.2 & 21.2 & 34.4 & & 29.1 & 33.2 & 42.8 & 67.8 & 96.6 \\
\midrule
Astrolabe gain & $+30.2\%$ & \textbf{+36.0\%} & $+33.8\%$ & $+11.9\%$ & $+8.1\%$ & & $+74.7\%$ & \textbf{+76.5\%} & $+53.2\%$ & $+16.2\%$ & $+17.2\%$ & & $+0.0\%$ & $+1.1\%$ & $+1.0\%$ & $+3.9\%$ & \textbf{+5.6\%} & & $-0.4\%$ & $+0.0\%$ & $-1.3\%$ & $+3.9\%$ & \textbf{+10.4\%} \\
\bottomrule
\end{tabular}%
}
\endgroup
\label{tab:request_metrics_snapshot}
\end{table*}

INFaaS++ performs well at low and moderate load but collapses as QPS grows: its memory-load heuristic sees only current memory and cannot anticipate incoming requests. Llumnix- does better at high load by correcting for known prefill demand, yet still misses decoding dynamics. Astrolabe's predictor simulates the full request lifecycle, routing around overloaded nodes before they collapse, for large latency and throughput gains without sacrificing capacity.

Across the 17-point sweep, Astrolabe's peak reductions over the best load-aware baseline are 38\% mean TTFT, 78\% TTFT-P99, 7\% mean E2E, and 12\% E2E-P99, with up to 5\% higher throughput; relative to the worst baseline, the gains reach 97\% lower TTFT and 53\% higher throughput. Astrolabe's per-request overhead stays under 3\% of E2E within capacity; its per-predictor CPU shows a $\sim$\cpuPoTwoVsFanoutRatio{}-fold reduction relative to full fanout (\S\ref{section:cpu_overhead}), which probes all instances before dispatching. Astrolabe's tail-TTFT gain over other schedulers follows a U-shaped trend across QPS (Figure~\ref{fig:request_metrics}, TTFT P99): gains are strongest at low load and overload, but Astrolabe falls behind Llumnix- from QPS 29 to 31. At low load the cluster is mostly idle, so the simulator predicts incoming-request latency accurately and Astrolabe routes around even transient stragglers. Idleness alone does not explain this advantage: Random and Round-Robin route without prediction yet trail Astrolabe's TTFT at the same low loads (Figure~\ref{fig:request_metrics}). At high load, herding on overloaded nodes becomes severe and is mitigated by Astrolabe's randomized dispatch and anticipation of request dynamics to avoid collapse and excessive preemptions.
At moderate load, the misprediction of request dynamics causes some suboptimal routing decisions that specifically affect tail latency, which makes the capacity gain less pronounced as judged by the TTFT-P99 metric (31.6 vs 31.5). However, the mean TTFT and E2E metrics are less sensitive to such tail events, so they still show consistent gains across the sweep. This suggests that even when the simulator's accuracy is not perfect, Astrolabe's overall scheduling decisions are still improved compared to non-predictive baselines, especially in terms of mean latency and throughput.

\begin{figure*}[!tb]
        \centering
        \includegraphics[width=0.93\linewidth]{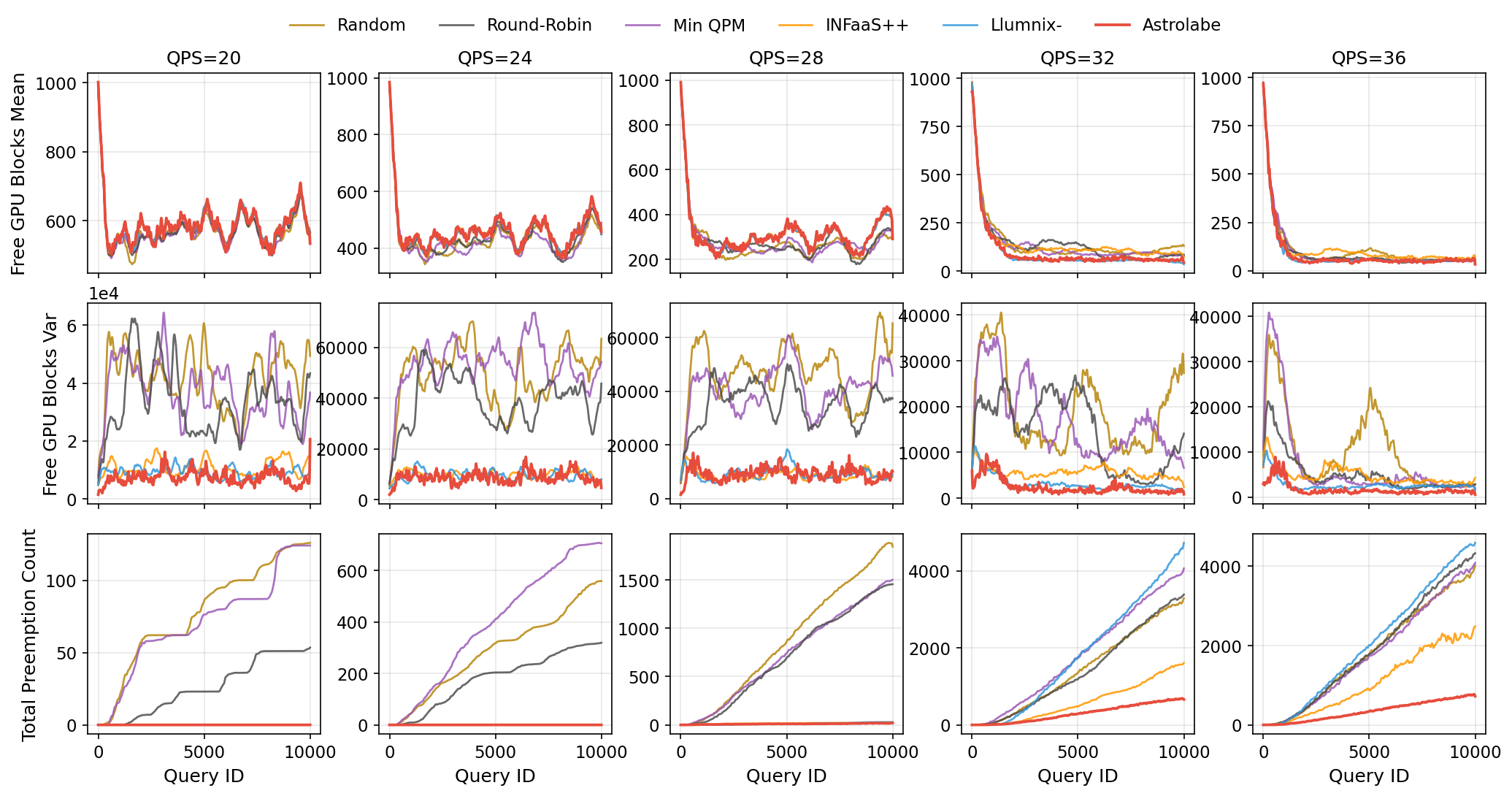}
         \caption{GPU memory blocks and preemption dynamics across QPS.}
         \Description{These time-related figures have been smoothed using a unified Gaussian filter to reduce noise and enhance readability.}
         \label{fig:resource}
\end{figure*}

Figure~\ref{fig:resource} also reports cross-instance variance of free GPU blocks and cumulative preemptions (both are Gaussian-smoothed with the same parameters to improve readability). Below capacity (QPS\,$\le$\,29) Astrolabe keeps variance low and preemptions near zero. The gap explodes at capacity: at QPS=32 Llumnix- accumulates $\sim$1.8k preemptions vs Astrolabe's $\sim$300 ($\sim$6$\times$ fewer); at overload (QPS 33--36) Llumnix- saturates around $\sim$2.1k preemptions while Astrolabe's count grows near-linearly and stays under 350. This aligns with the sustained TTFT/E2E lead past capacity.

\subsection{End-to-end Comparison with Llumnix}\label{section:llumnix_comparison}

We compare Astrolabe with full Llumnix~\cite{sun2024llumnix}, which performs dynamic load rebalancing via live KV-cache migration. Our A30 cluster (single-GPU-per-host, 25\,Gbps Ethernet) is unsuitable for migration due to network constraints. We therefore use a 2-node cluster with 4$\times$ A100-40GB GPUs per node, analogous to Llumnix's original 4-node A10 testbed. Llumnix moves KV-cache via Ray RPC by default and does not support chunked prefill (CP) in the evaluated configuration. To keep the comparison faithful to the released system, we keep Llumnix's default Ray-RPC migration path and do not retune its migration policy for our workload; we change only deployment parameters needed to run the same model and instance count on the A100 cluster. Symmetrically, Astrolabe also runs untuned, with the same defaults ($k{=}2$, 10-token extension) used throughout \S\ref{section:evaluation}. To isolate the scheduler's contribution from chunked-prefill backend effects, we evaluate Astrolabe with and without CP against Llumnix across an identical QPS sweep. With larger A100 memory (40\,GB vs 24\,GB), we raise the batch size to 96. All configurations serve Llama-2-7B across 8 instances (4 per node, TP=1 for each), using FlashInfer as the backend.

\begin{figure}[!tb]
    \centering
    \includegraphics[width=1.0\linewidth]{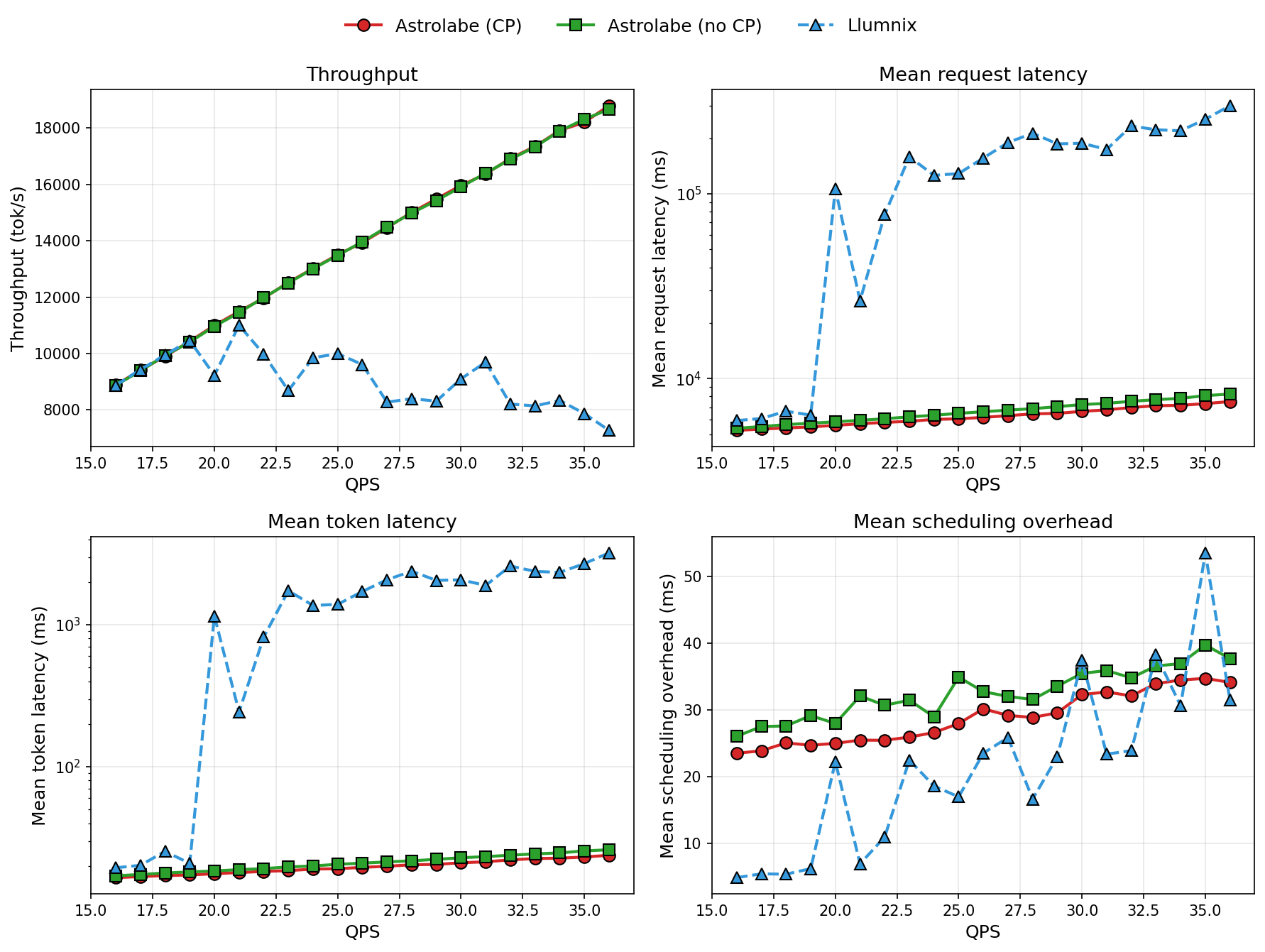}
    \caption{Astrolabe with and without chunked prefill vs full Llumnix on the A100 cluster}
    \Description{Throughput, mean request latency, mean token latency, and scheduling overhead versus QPS for Astrolabe with and without chunked prefill and full Llumnix on the A100 cluster.}
    \label{fig:llumnix_comparison}
\end{figure}

Figure~\ref{fig:llumnix_comparison} reports throughput, mean request latency, token latency, and scheduling overhead. Astrolabe's mean scheduling overhead is higher than Llumnix's at low load (23--29\,ms vs 5--7\,ms at QPS\,$\le$\,19) due to pre-dispatch predictor queries and simulation, but stays within tens of milliseconds. Through QPS=19, both systems serve comparably; CP gives Astrolabe a small edge (5.5\,s vs 6.3\,s mean E2E at QPS=19), while the CP and no-CP Astrolabe curves remain within $\sim$10\% of each other. The high-load gap is therefore not explained by CP alone: Astrolabe without CP already outperforms Llumnix substantially. Llumnix collapses at QPS=20, where mean request latency jumps from 6.3\,s to 106\,s and mean token latency from 21\,ms to 1.15\,s. This is consistent with migration-induced PCIe/NIC contention: repeated KV-cache transfers compete with serving traffic until migration overhead dominates E2E latency. At QPS=36, the direct no-CP comparison already favors Astrolabe: Astrolabe without CP delivers 2.57$\times$ the throughput (18.7k vs 7.3k\,tok/s), 36$\times$ lower mean request latency (8.2\,s vs 300\,s), and 123$\times$ lower mean token latency (26\,ms vs 3.2\,s) than Llumnix. Enabling CP widens these to 2.58$\times$/40$\times$/135$\times$.

\subsection{\texorpdfstring{$k$-tunable Dispatch: from Po2 to Fanout}{k-tunable Dispatch: from Po2 to Fanout}}\label{section:po2_ablation}

Astrolabe's core dispatch policy is \emph{power-of-$k$ choices}: for each request, sample $k$ random instances, query each Predictor in parallel, and route to the one with the lowest predicted latency. Increasing $k$ trades higher control-plane fanout cost (linear in $k$) for better load-balancing quality and stronger mitigation of herding~\cite{mitzenmacher2001,10.1137/S0097539795288490}. We evaluate $k \in \{2, 4, 8, 12\}$; $k{=}12$ recovers full fanout.

\begin{figure}[!tb]
    \centering
    \includegraphics[width=1.0\linewidth]{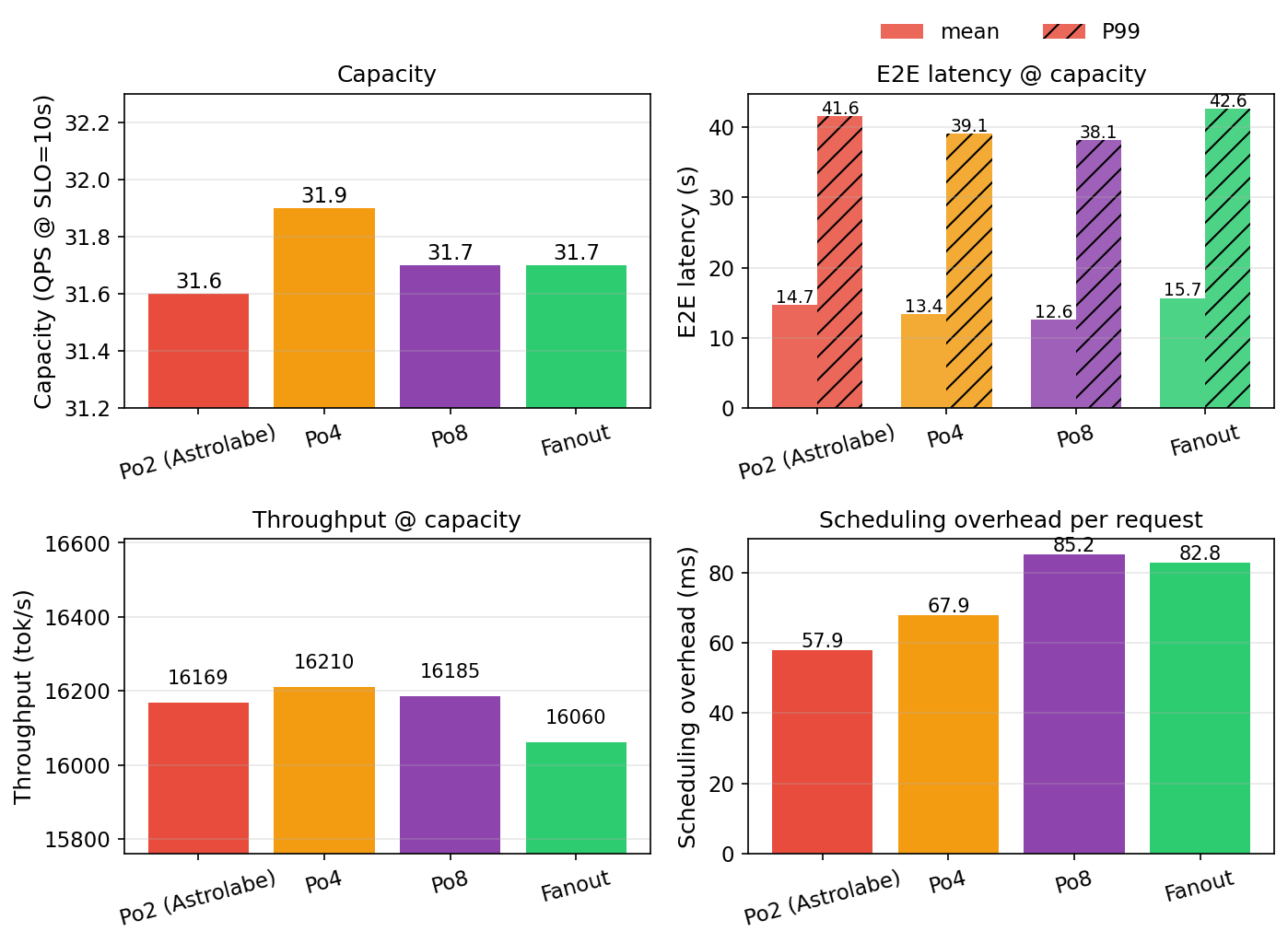}
    \caption{Power-of-$k$ choices ablation}
    \Description{Capacity, E2E latency, throughput, and scheduling overhead for power-of-k choices variants from k=2 to full fanout.}
    \label{fig:po2_comparison}
\end{figure}

Figure~\ref{fig:po2_comparison} reports capacity, mean/P99 E2E latency at each scheduler's own capacity, throughput, and per-request scheduling overhead. Capacity is effectively tied from $k{=}2$ upward: Po2-est (\emph{-est}: RoBERTa-estimated lengths, the default; \emph{-oracle}: ground-truth lengths) reaches \capPoTwoEst{} QPS vs Po4-est \capPoFourEst{} (the maximum), Po8-est \capPoEightEst{}, and Fanout-est \capFanoutEst{}, all within a $\pm$0.3\,QPS band at SLO\,=\,\sloSec{}s. Scheduling overhead grows with $k$ through $k{=}8$ and then plateaus, because each dispatch fans out to more predictors and is bounded by the slowest reply.

Randomized probing also mitigates herding: when bursty requests arrive within one round trip, deterministic Fanout can compute identical predictions and overload the same instance, while power-of-$k$ randomizes each request's candidate set. This helps the sampled variants keep lower E2E latency and higher throughput than Fanout in Figure~\ref{fig:po2_comparison}. In this sweep, $k{=}4$ gives the best empirical tradeoff: it has the highest capacity and throughput, while retaining the second-lowest E2E latency and scheduling overhead. We nevertheless use Po2 as the default in the main evaluation to follow the standard power-of-two-choices convention without tuning $k$ on the evaluation workload.

\subsection{Oracle-Length Upper Bounds}\label{section:oracle_bounds}

Astrolabe uses a RoBERTa length estimator (MAPE 24.4\%, \S\ref{section:predicting accuracy}). Replacing it with ground-truth (``oracle'') lengths gives an upper bound on the gains Po2 and Fanout can obtain from a perfect length predictor. Length-controllable workloads do exist, such as constrained decoding~\cite{geng2025jsonschemabench,beurerkellner2022lmql,willard2023efficient} and length-adaptive applications~\cite{reddy2025bellman}, which can provide the exact length of each request at dispatch time. In addition, applications with standard output formats can have highly deterministic output lengths even without model-based prediction, making the oracle mode realistic for these scenarios.
\begin{figure}[!tb]
    \centering
    \includegraphics[width=1.0\linewidth]{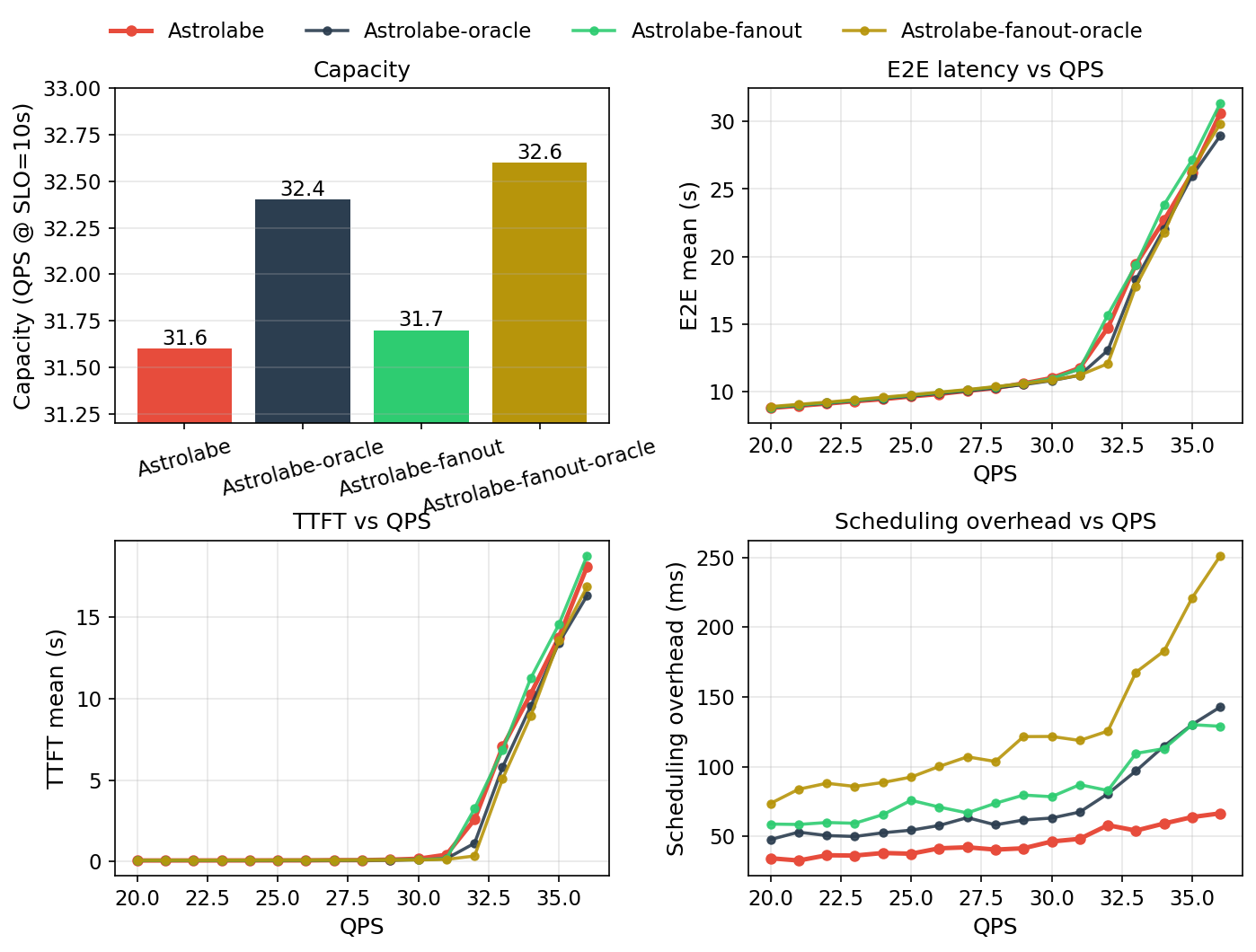}
    \caption{Oracle-length upper bounds}
    \Description{Capacity, mean E2E latency, mean TTFT, and scheduling overhead comparing estimated and oracle lengths for Po2 and Fanout.}
    \label{fig:oracle_comparison}
\end{figure}

Figure~\ref{fig:oracle_comparison} compares the four variants across capacity, mean E2E latency, mean TTFT, and scheduling overhead. Oracle adds $\sim$0.8\,QPS for Po2 (\capPoTwoEst{}\,$\to$\,\capPoTwoOracle{}) and $\sim$0.9\,QPS for Fanout (\capFanoutEst{}\,$\to$\,\capFanoutOracle{}) at SLO\,=\,\sloSec{}s, a $\sim$3\% capacity gain. The default Astrolabe Po2 variant has the lowest scheduling overhead, while Fanout-oracle has the highest. One candidate explanation is that estimated lengths are more uniform than actual ones, increasing the simulator's cache hit rate for repeated batch-latency lookups; oracle variants expose more length diversity and thus higher simulation overhead. Validating this hypothesis---e.g., by bucketing lengths or replaying a synthetic trace whose lengths match predictions exactly---is a targeted robustness check we leave to future work. Even for Fanout-oracle, however, scheduling overhead stays below 3\% of total E2E latency in this sweep and tends to decrease once capacity is reached, as E2E latency then spikes more rapidly.

\subsection{Ablation Study Under Bursty Workloads}\label{section:burstiness}
Real-world LLM traffic exhibits bursty patterns~\cite{burstgpt}. We vary burstiness with gamma-distributed arrivals and set the shape parameter $\alpha$ to 0.25, 0.5, 1.0, and 2.0. Lower $\alpha$ produces burstier, higher-variance inter-arrival times; at $\alpha{=}1.0$, the arrival process reduces to the Poisson case with exponential inter-arrival times. All experiments run at QPS=32 on the A30 cluster against Llumnix-, and we record E2E latencies; each ($\alpha$, scheduler) point is a single 10k-request run, with the cross-deployment reproducibility reported in \S\ref{section:request_performance}.

\begin{figure}[!tb]
    \centering
    \includegraphics[width=1.0\linewidth]{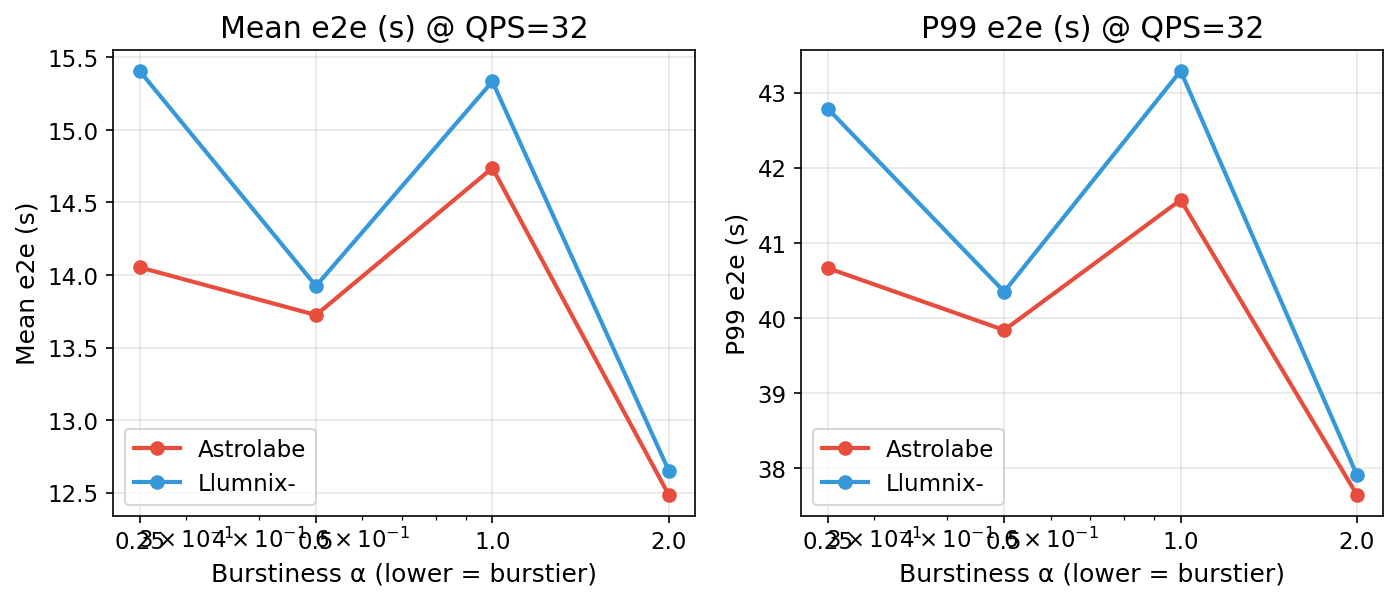}
    \caption{Latency under bursty workloads}
    \Description{E2E latency under gamma arrival processes with varying burstiness for Astrolabe and Llumnix-.}
    \label{fig:burstiness}
\end{figure}

As seen in Figure~\ref{fig:burstiness}, Astrolabe matches or beats Llumnix- across all four levels. Under high burstiness ($\alpha{=}0.25$), Astrolabe's mean E2E latency is about 9\% lower than Llumnix-; under near-regular arrivals ($\alpha{=}2.0$), the gap narrows but Astrolabe retains a small edge. This pattern is consistent with randomized probing decorrelating placement decisions under bursts, reducing request herding while keeping per-request probing cost fixed at two predictors. Moving from $\alpha{=}2.0$ to $\alpha{=}0.25$ raises mean E2E by $\sim$13\% for Astrolabe vs $\sim$22\% for Llumnix-.

\subsection{Sensitivity to Prediction Errors}\label{section:error_sensitivity}

A key design question is whether length-prediction error (24.4\%, Table~\ref{table:length_prediction}) and simulator error (10--15\%, Figure~\ref{fig:latency_prediction}) undermine dispatch quality. We answer this by injecting noise well beyond both: $v' = v \cdot (1 + \epsilon)$ with $\epsilon \sim \mathcal{N}(0, \sigma)$, clamped to $[0.1v, 5.0v]$ to avoid invalid values (e.g., negative lengths). We inject noise independently into the length estimate and each instance's simulated latency. We sweep $\sigma$ over 0\%, 25\%, 50\%, and 100\% per axis, producing a 4$\times$4 grid at QPS=32 with 10k requests per cell, covering error profiles up to noise-dominated extremes.

\begin{figure}[!tb]
    \centering
    \includegraphics[width=1.0\linewidth]{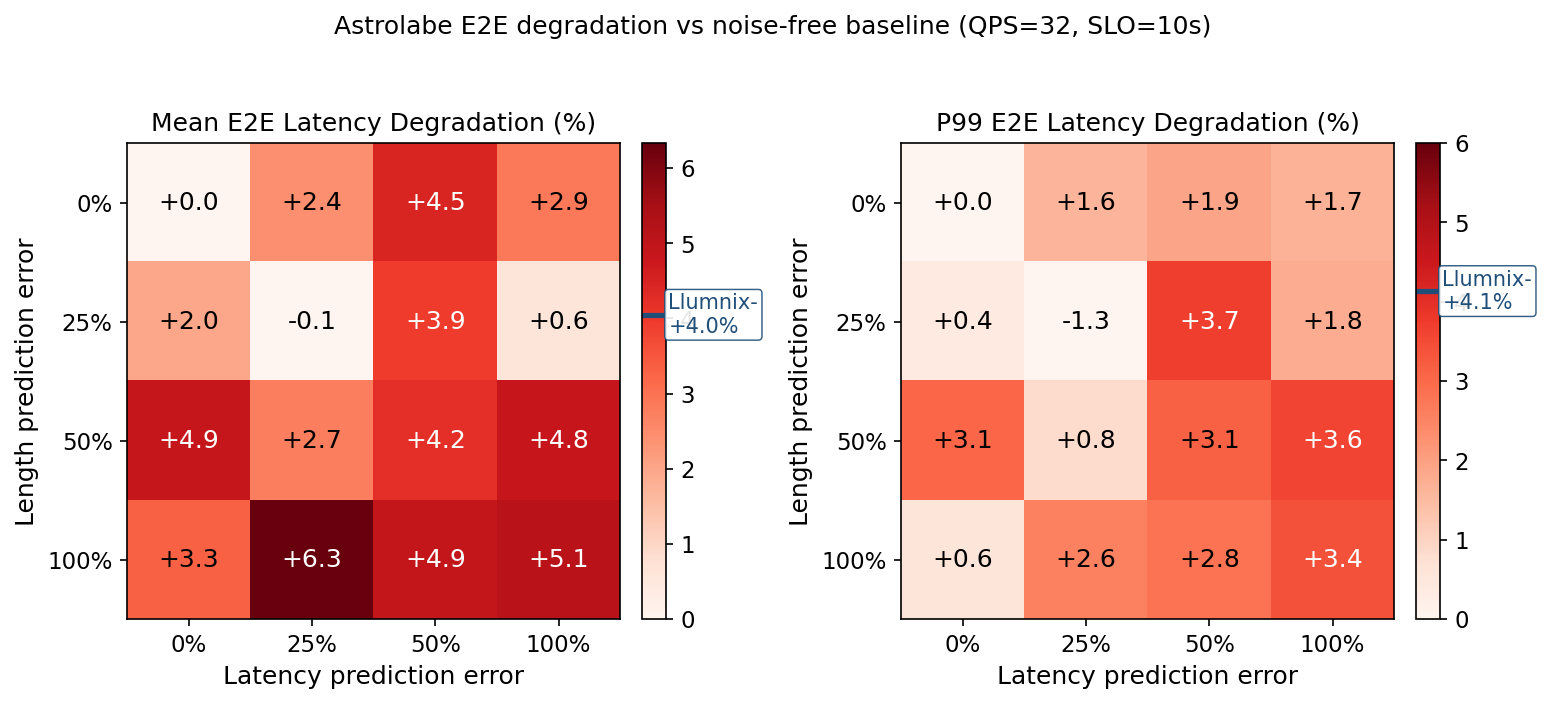}
    \caption{Sensitivity of Latency to Prediction Error}
    \Description{Heatmaps of mean and P99 E2E degradation as length-estimation and latency-prediction noise increase.}
    \label{fig:error_heatmap}
\end{figure}

Figure~\ref{fig:error_heatmap} plots mean (left) and P99 (right) E2E degradation vs the noise-free baseline. The largest mean degradation is +6.3\% at (100\%,\,25\%), while the extreme (100\%,\,100\%) cell is +5.1\%; most cells remain within the $\pm$4\% single-deploy noise floor. Figure~\ref{fig:error_heatmap} overlays Llumnix-'s degradation (+4.0\% mean, +4.1\% P99) as a colorbar reference. Astrolabe stays below this line at moderate noise; at high length-error settings, several mean-latency cells exceed it modestly, but P99 degradation remains below Llumnix- in every cell (maximum +3.7\%). Thus predictor-guided dispatch retains value across the injected-noise range. Three factors keep Po2 robust: (i) dispatch uses \emph{relative} ranking between the two sampled probes, so common-mode length error cancels and only large per-instance latency noise can flip a ranking; (ii) underlying latency gaps between loaded and idle instances are usually 2--10\,s, so moderate noise rarely reverses orderings; (iii) Po2's random probe set prevents a bad ranking from repeatedly steering requests to the same overloaded instance, the same randomization principle as in~\S\ref{section:po2_ablation}. Length underestimation is also mitigated by the auto-extension (\S\ref{section:model_instance}), which applies the same 10-token extension rule across candidate simulations. Prediction quality can also drift as workloads shift. Since Astrolabe tolerates errors well beyond the measured 24.4\% MAPE and the length model is cheap to re-fine-tune offline, periodic retraining could be employed to track such drift and maintain performance gains.

\subsection{CPU Overhead: Po2 vs Fanout}\label{section:cpu_overhead}

Astrolabe's Predictor runs CPU-side simulation off the GPU-bound inference path. The control-plane cost is governed by the number of predictor probes per dispatch: Fanout queries all 12 predictors, while Astrolabe's default Po2 policy queries only two. We measure per-predictor CPU utilization and RSS memory on the A30 cluster.
\begin{figure}[!tb]
    \centering
    \includegraphics[width=1.0\linewidth]{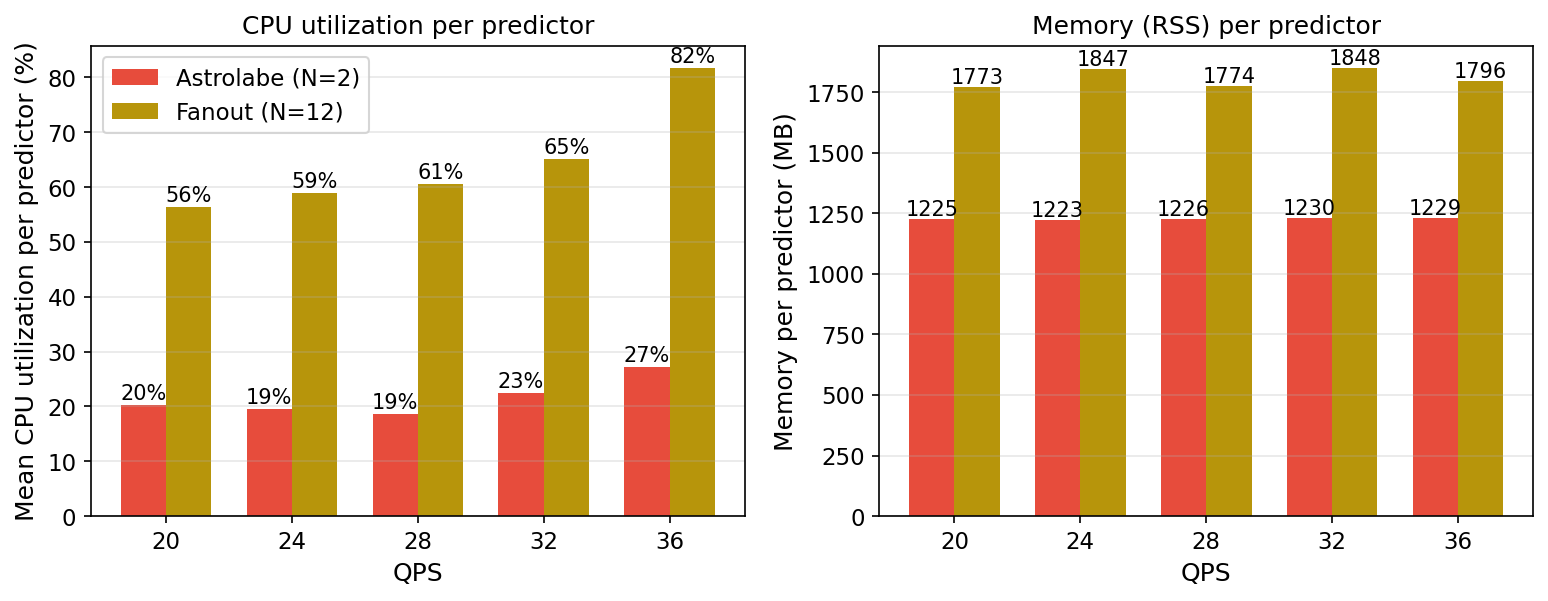}
    \caption{Per-predictor CPU utilization and memory across QPS}
    \Description{Bar charts of mean CPU utilization and RSS memory per predictor across QPS for Po2 and Fanout.}
    \label{fig:cpu_overhead}
\end{figure}
As seen in Figure~\ref{fig:cpu_overhead}, at QPS=20 mean per-predictor CPU is \cpuPoTwoMeanAtTwenty{}\% for Po2 and \cpuFanoutMeanAtTwenty{}\% for Fanout, a \cpuPoTwoVsFanoutRatio{}-fold reduction. The gap persists across QPS 20, 24, 28, 32, and 36 because Fanout sends every request to all 12 predictors, whereas Po2 sends each request to only two. Per-predictor RSS, in contrast, is stable across QPS (Po2 1225\,MB, Fanout 1773\,MB at QPS=20). The scaling difference is thus CPU-bound: RSS is a fixed cost, while Fanout's aggregate predictor CPU grows with cluster size as each request is sent to all predictors; Po2 keeps per-request fanout constant at two probes.

\subsection{Generality Study}\label{section:general_study}
\begin{table}[tb]
\caption{Capacity across workload/configuration changes. bs/cs denotes batch size and chunk size.}
\setlength{\tabcolsep}{4pt}
\resizebox{\linewidth}{!}{%
\begin{tabular}{lcccc}
\toprule
Scheduler    & bs=24 & cs=2048 & Qwen2-7B & BurstGPT \\
\midrule
Astrolabe-oracle  & 29.1 & 31.9 & \capQwenPoTwoOracle{} & \capBurstGPTPoTwoOracle{} \\
Astrolabe         & 28.5 & 31.6 & \capQwenPoTwoEst{}    & --  \\
Llumnix-          & 26.6 & 31.3 & \capQwenLlumnix{}     & 61.6 \\
Gain (est/oracle) & 7.1\%/9.4\% & 1.0\%/1.9\% & 6.0\%/12.6\% & --/4.7\% \\
\bottomrule
\end{tabular}%
}
\label{tab:scheduler_perf_scaled}
\end{table}

LLM serving settings change dynamically (batch size, chunk size, model, dataset); heuristic static rules ignore such changes and drift (\S\ref{section:scheduling_issues}), while Astrolabe's predictor simulates the specific backend configuration and absorbs them automatically. Table~\ref{tab:scheduler_perf_scaled} compares Astrolabe against Llumnix- across four variations.

Across these variations: (1)~\emph{batch/chunk}---with batch size reduced 48$\to$24, Astrolabe reaches 28.5 QPS vs Llumnix- 26.6 QPS (+7.1\%) and Astrolabe-oracle reaches 29.1 QPS (+9.4\%); with chunk size raised 512$\to$2048, the gap narrows to 1.0\%/1.9\%, similar to the main-experiment gap, since a larger chunked-prefill step perturbs the predictor's inputs less than a batch-size change. (2)~\emph{model: Qwen2-7B}---after retraining the RoBERTa length model on ShareGPT tokenized with Qwen, Astrolabe reaches \capQwenPoTwoEst{}~QPS vs Llumnix- \capQwenLlumnix{}~QPS (+6.0\%, +12.6\% with oracle lengths). (3)~\emph{dataset: BurstGPT}~\cite{burstgpt}---the trace lacks prompt content, preventing length-predictor retraining,
so we compare only oracle variants: Astrolabe-oracle reaches \capBurstGPTPoTwoOracle{}~QPS vs Llumnix- 61.6~QPS (+4.7\%). Across all four, Astrolabe or its oracle variant matches or outperforms Llumnix-, confirming robustness to backend, model, and workload changes.

\section{Conclusion}
Astrolabe combines response-length estimation, simulation-based latency prediction, and randomized dispatch for multi-instance LLM serving. Across our experiments, prediction-guided one-shot dispatch matches or outperforms migration-based rebalancing while avoiding KV-cache transfer overhead. Astrolabe treats each instance as one endpoint regardless of intra-instance parallelism (validated at Llama-2-70B TP=4, \S\ref{section:predicting accuracy}). We leave larger-scale multi-GPU evaluation, richer failure handling, and length-model retraining under workload drift as future work.

\begin{acks}
We thank the anonymous SYSTOR reviewers and our shepherd for their constructive feedback, and CloudLab~\cite{cloudlab} for providing the computing resources used in our experiments.
\end{acks}

\bibliographystyle{ACM-Reference-Format}
\bibliography{reference}

\end{document}